\documentclass[prb,twocolumn,floatfix,showpacs]{revtex4}
\pdfoutput=1
\usepackage{amsfonts}
\usepackage{amssymb,amsmath}	
\usepackage{bm}
\usepackage{graphicx}
\usepackage{hyperref}
\begin{document}
\title{Competing order, Fermi surface reconstruction,  and quantum oscillations   in underdoped high temperature superconductors}
\author{Ivailo Dimov}
\author{Pallab Goswami}
\author{Xun Jia}
\author{Sudip Chakravarty}
\affiliation{Department of Physics and Astronomy, University of California at Los Angeles,
Los Angeles, California 90095, USA}
\date{\today }
\begin{abstract}
We consider incommensurate  order parameters for electrons
on a square lattice which reduce to $d$-density wave  order when the
ordering wavevector ${\bf Q}$ is close to ${\bf Q}_0 = (\pi/a,\pi/a)$, $a$ being the lattice spacing,  and describe
the associated charge and current distributions within a single-harmonic approximation
that conserves current to lowest order. Such incommensurate  orders can arise at the
mean-field level in extended Hubbard models, but the main goal here is to explore thoroughly the consequences within a Hartree-Fock approximation. 
We find that Fermi surface reconstruction in the underdoped regime can correctly capture the phenomenology of the recent quantum oscillation experiments that suggest incommensurate order, in particular the de Haas-van Alphen oscillations of the magnetization in high fields and very low temperatures in presumably the mixed state of these superconductors. For  10\% hole doping  in $\mathrm{YBa_{2}Cu_{3}O_{6+\delta}}$, we find in addition to the main frequency around 530 T arising from the electron pocket and a hole frequency at around 1650 T,  a new low frequency from a smaller hole pocket at 250 T for which there are some indications that require further investigations. The oscillation corresponding to the electron pocket will be further split due to bilayer coupling but the splitting is sufficiently small to require more refined measurements. The truly incommensurate $d$-density wave breaks both time reversal and inversion but the product of these two symmetry operations is preserved. The resulting Fermi surface splits into spin up and spin down sectors that are inversion conjugates. Each of the spin sectors results in a band structure that violates reflection symmetry, which can be determined in spin and angle resolved photoemission spectroscopy. For those experiments such as the current photoemission experiments or the quantum oscillation measurements that cannot resolve the spin components the bands will appear to be symmetric because of the equal mixture of the two spin sectors. There is some similarity of our results with the spiral spin density wave order which, as pointed out by Overhauser, also breaks time reversal and inversion. Calculations corresponding to higher order commensuration produces results similar to anti-phase spin stripes, but appear to us to be an unlikely explanation of the experiments. The analysis of the Gorkov equation in the mixed state shows that the oscillation frequencies are unshifted from the putative normal state and the additional Dingle factor arising from the presence of the mixed state can provide a subtle distinction between the spiral spin density wave and the $d$-density wave, although this is very difficult to establish precisely. 
\end{abstract}

\maketitle

\section{Introduction}\label{intro}

Despite more than twenty years of intense effort,  the telltale evidence of competing order parameters in the phase diagram
of high temperature cuprate superconductors remains an enigma. Other than antiferromagnetism of the undoped materials,~\cite{Chakravarty:1989} and stripe order 
at special doping and special materials,~\cite{Kivelson:2003} not much is truly understood. Yet, the attractive option
of classifying the pseudogap as a broken symmetry state has spurred intense interest.~\cite{Zhang:1997,Kivelson:1998,Varma:1999,Caprara:1999,Chakravarty:2001} Indeed,
this option has the great power to unify   the diverse phenomenology of these materials. We are strongly motivated 
by the recent quantum oscillation measurements~\cite{Doiron-Leyraud:2007,LeBoeuf:2007,Yelland:2008,Bangura:2008,Jaudet:2008,Sebastian:2008} to address this question anew. These are the Shubnikov-de Haas (SdH) effect, the de Haas-van Alphen (dHvA)effect  and the oscillations of the Hall coefficient ($R_{H}$). The rapidly evolving experimental situation  
suggests,  under various circumstances, evidence  of commensurate and incommensurate order. It is perhaps best at this time to explore as thoroughly as possible both kinds of order within a very general framework.  Such a framework is provided by the  Hartree-Fock theory. 

Our basic assumption,  justified when compared to experiments, is that the observed quantum oscillations
in high magnetic fields and low temperatures in underdoped $\mathrm{YBa_{2}Cu_{3}O_{6+\delta}}$ arise from the Landau levels of the quasiparticles  of the Fermi liquid  corresponding to the putative normal state. It is known, and shown here, that in the mixed state of many superconductors the oscillations remain unshifted in their  frequencies  from the normal state but exhibit increased damping. Thus, the beauty of the present experiments is that one can in principle determine 
the closed Fermi surfaces, through the Onsager relation  that involves only the fundamental constants and the Fermi surface area,  of the normal state without having to cross the  upper critical field.  At  face value these observations have revealed some striking facts not noticed previously.~\cite{Chakravarty:2008} The most important of which is the existence of both positively and negatively charged quasiparticles in a hole doped cuprate at variance with the conventional wisdom. That these particles behave like quasiparticles in a Fermi liquid, even  in the underdoped regime, which has so far  been plagued with complex theoretical concerns involving the proximity to a Mott insulator,~\cite{Lee:2008} is striking to say the least and should be of great value in uncovering the mystery of these enigmatic superconductors. One of us has elaborated on the striking nature of these discoveries and has explained that the root of these observations is the Fermi surface reconstruction due to broken symmetries.~\cite{Chakravarty:2008} One of us has also elucidated  the role of commensurate singlet $d$-density wave (DDW) to explain the oscillations of the Hall coefficient.~\cite{Chakravarty:2008b} The Hall measurement is more telling than either SdH or dHvA because it contains the striking information regarding the sign of the charge carriers. There is evidence of the presence of electron pockets in what is really a hole doped material, which is difficult to explain  without  spontaneously breaking symmetries in the putative normal state, in particular by the featureless liquid of  resonating valence bond theories.~\cite{Anderson:1987}  

For a particle-hole condensate, which DDW is, the symmetry of the orbital wave function  does not constrain the symmetry of the spin wave function.~\cite{Halperin:1968}  Triplet DDW has not yet been adequately explored  and remains a subject for future work. The  more recent experiments have, however, revealed some indication of an incommensurate ordering,~\cite{Sebastian:2008} which was only briefly touched upon previously.~\cite{Chakravarty:2008b,Jia:2008} Here we focus on the incommensurate case. 

Before we begin, it is important to state what a Hartree-Fock theory can or cannot accomplish. A Hartree-Fock theory can identify the possible broken symmetries and phases, but cannot establish that such symmetries are indeed broken. The prime deficiency  is the absence of all fluctuation effects, but often such deficiencies can be remedied by educated guesses. Of course, deep inside a broken symmetry phase critical fluctuations are absent and some of the collective modes can be physically identified from the symmetries of the order parameters. Most importantly, symmetries protect the Hartree-Fock calculations; those properties that are determined by the symmetries alone can be determined in the weak interaction limit, where the calculations are better controlled, from a suitable effective Hamiltonian. The results protected by symmetries should be at least qualitatively valid even in the strong interaction limit.~\cite{Kopp:2007} However, the thermal properties in low dimensional systems are badly predicted by the Hartree-Fock theories, because fluctuations are often very important. Nonetheless, it is expected that the results at zero temperature retain a considerable degree of validity, except close to quantum critical points where quantum fluctuations become important. 

When a commensurate one-dimensional ($1 D$) charge density wave (CDW) is doped,
the resulting charges may be viewed as defects in the CDW.
If the interactions between them are sufficiently strong
compared to their effective kinetic energy, then they will form an ordered lattice
(which can be stabilized by a crossover to three dimensions if there are
many $1D$ chains coupled together)
of defects and the CDW will become incommensurate with the lattice.
This can be energetically more  favorable than simply doping holes into the rigid band
formed in the presence of CDW order at a fixed wavevector. The reason,
in the latter case, is the single-particle energy gap of the CDW.
In a two-dimensional DDW
state, however, there are nodes in the order parameter, so it may be more favorable
for doped holes to simply go to the nodes and expand into Fermi pockets
(in quasi-$1D$ systems, however, this might not be possible, so doping must
lead to incommensurability~\cite{Schollwoeck:2003}).
However, this cannot persist indefinitely if the DDW state is stabilized by
approximate nesting, since the Fermi surface  would eventually move away
from the nesting wavevector. Thus, incommensurate DDW order
is a possibility, at least over a range of doping. In this paper,
we explore the phenomenology of this possibility. But it is very difficult to determine if this incommensuration happens before the commensurate gap collapses and some other order takes over.
In this respect the tendency of the spin density wave (SDW) to incommensurate is
stronger because it is gapped everywhere on the Fermi surface, being intrinsically a $s$-wave
object.

Near half-filling, the Hubbard model, the $t-J$ model, and generalizations
of these appear to have many phases which are close in
energy.
Thus, small perturbations can strongly influence the competition between them.
Consequently, it should not be too surprising that
experiments on the cuprate superconductors have also uncovered some evidence
for a cornucopia of phases, particularly on the underdoped side 
of the phase diagram, which appear in particular materials and
for certain values of the doping level, temperature, magnetic
field, etc.
Depending on the interfacial energies between these phases,
one way in which their competition can be resolved
is through the formation of stripes~\cite{Emery:1999}
or other inhomogeneous patterns of microscale phase-separation.
Furthermore, as the cuprates are doped, their Fermi surfaces
evolve away from the two-fold commensurate nesting wavevector ${\bf Q}_{0}=(\pi/a,\pi/a)$,
so we would expect translational symmetry-breaking order
parameters to occur at incommensurate wavevectors.
Hence, it would seem
important for any theory of the pseudogap to
incorporate the tendency towards incommenuration,
which we take as further impetus to study
incommensurate  order parameters
related to DDW order,  the IC-DDW.
Direct attempts to measure DDW order through
neutron scattering have neither ruled it out ~\cite{Stock:2002,Fauque:2006} 
nor unambiguously verified its presence, but there are 
intriguing suggestions that it may be present~\cite{Mook:2002,Mook:2004}.

In Sec. II the construction of both commensurate and incommensurate order parameters are discussed.
This discussion serves to motivate our principal approximations. Sec. III on fermiology is the heart of our 
paper. We begin with the calculation of the electronic spectra of a chosen IC-DDW order and deduce  the corresponding 
areas of Fermi pockets that, through the Onsager relation, lead to the oscillation frequencies. The similarity of these results 
with the spiral SDW scenario is explained. Higher order commensurate DDW results are also discussed. These also turn
out to be similar to the anti-phase spin stripe calculations.~\cite{Millis:2007} Thus, simply from a Hartree-Fock calculation it is not possible to 
single out a mechanism. One must invoke other considerations to distinguish between mechanisms. Next we show that interlayer coupling can lead to bilayer splitting of the areas of the electron pockets leaving the hole pockets essentially untouched. The subtleties of this result are discussed in some depth. Sec. IV contains a calculation of the self energy of the quasiparticles in the mixed state patterned after the work of Stephen,~\cite{Stephen:1992} which we find to be by far the soundest approach, despite its deficiencies. The calculation for the nodal quasiparticles of DDW is a little subtle and leads to a vortex contribution of the scattering rate that we find very reasonable. If this calculation can be better controlled, it may be possible to distinguish between the SDW and the DDW scenarios. Given the approximate nature of this calculation, which can only yield a semiquantitative estimate, it seemed sufficient to illustrate our calculation with the commensurate case. Of course, the vortex contribution to the scattering rate is only a part of the total that also contains an impurity contribution.  In the concluding section, Sec. V, we discuss our thoughts on microscopic models and whether or not  incommensuration does actually occur.  We also elaborate on experimental consequences of our work and point out the unresolved issues and future directions of research.

\section{Order Parameters}
\subsection{Commensurate}
Before addressing the IC-DDW order parameter, let us summarize some of the features of 
the  commensurate case.~\cite{Marston:1989,Nayak:2000} A commensurate singlet particle-hole condensate is defined 
by the order parameter
\begin{equation}\label{corder_k}
\langle{c}^{\dagger}_{\alpha'\bf k'}c^{}_{\alpha,{\bf k}}\rangle
=\pm iW_{\bf k}\,\delta_{\alpha',\alpha}
\delta_{{\bf k'},{\bf k+Q_0}},
\end{equation}
where ${\bf Q}_{0}= (\pi/a,\pi/a)$. Although the right hand side should correctly contain a  factor   $\pm iW_{\bf k}$, we will choose $iW_{\bf k}$ for clarity, as no confusion should arise. The electron destruction  operator, $c _{{\bf k}\alpha}$, is indexed by the wave vector $\bf k$ and spin   $\alpha$; $a$ is the lattice spacing. 
We say that the order is commensurate, or strictly $2$-fold commensurate, because $2{\bf Q_0}$ is a reciprocal lattice vector. 
The form factor $W_{\bf k}$ transforms non-trivially under the point group of the 
two dimensional square  lattice, defined by ${\bf R}= m a \hat{\bf x}+n  a \hat{\bf y}$, where $(m,n)$ is a set of integers positive or negative. It is related to the orbital angular momentum of the 
condensate. For DDW, 
\begin{equation}
W_{\bf k}=\frac{W_{0}}{2}(\cos{k_x}a-\cos{k_y}a),
\end{equation}
which corresponds to angular momentum  $\ell=2$ and
$d_{x^2-y^2}$ symmetry.  For a particle-hole condensate, the symmetry of the orbital wave function does not constrain the
spin wave function. Thus, there can also be a triplet DDW which corresponds to circulating staggered spin currents,~\cite{Nersesyan:1991} as opposed to charge currents. If we set $W({\bf k})$ by a function that transforms as the identity  and remove the factor $i$, the order parameter would correspond to the conventional CDW. Similarly, the triplet DDW is the $\ell =2$ generalization  of the conventional spin density wave (SDW) and is given by
\begin{equation}
\langle{c}^{\dagger}_{\alpha'\bf k'}c^{}_{\alpha,{\bf k}}\rangle
=iW_{\bf k}\, \hat{\bf n}\cdot {\boldsymbol \sigma}_{\alpha,\alpha'}
\delta_{{\bf k'},{\bf k+Q_0}},
\end{equation}
where $\boldsymbol \sigma$'s represent the Pauli matrices, and $\hat{\bf n}$ is a unit vector in the spin space. Once again, if we replace  $iW_{\bf k}$ by unity, this order parameter is the conventional SDW.

\begin{figure}[htb]
\centerline{\includegraphics[height=3.0in]{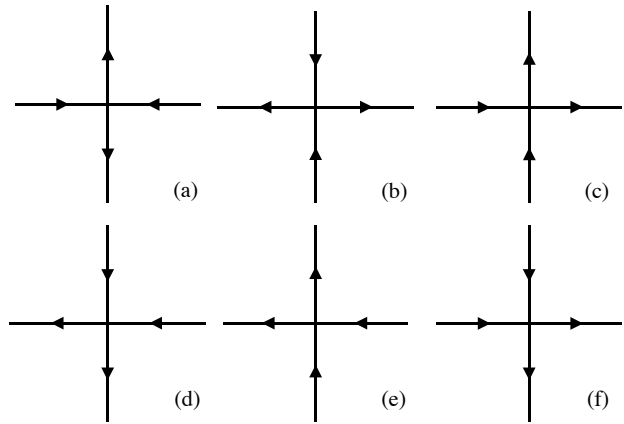}}
\caption{The 6-vertex model generalization of DDW. The vertices (c)-(f) are the singlet $p$-wave density wave vertices. Such a model has a thermodynamic  phase transition to the low temperature DDW state but with no specific heat anomaly.} 
\label{fig:6vert}
\end{figure}

Because $W_0$ is real, the order in Eq.~\eqref{corder_k}
breaks time-reversal and hence represents a state with non-zero orbital 
current configuration. This is most easily seen in real space, where 
Eq.~\eqref{corder_k} can be written as (the spin indices $\alpha$ suppressed
for clarity):
\begin{equation}\label{corder_x}
\langle{c}^{\dagger}_{\bf R'}c_{\bf R}\rangle
=\pm  i(-1)^{m'+n'}\frac{W_0}{2}{\rm V}_{\bf R',R}  
\end{equation}
The $d$-wave hopping amplitude factor ${\rm V}_{\bf R',R}$ is
\begin{equation}\label{vertices}
{\rm V}_{\bf R',R} = -\delta_{{\bf R}',{\bf R}+\hat{\bf x}a}-\delta_{{\bf R}',{\bf R}-\hat{\bf x}a}+
	\delta_{{\bf R}',{\bf R}+\hat{\bf y}a}+\delta_{{\bf R}',{\bf R}-\hat{\bf y}a}
\end{equation}
and also represents a current-conserving vertex centered at ${\bf R}$ as
in Fig.\ref{fig:6vert} (a). 
A positive/negative sign in front of each term 
in Eq.~\eqref{vertices} corresponds to current going out/in of the vertex.
This analogy motivated a 6-vertex description of the transition to a 
commensurate DDW state,~\cite{Chakravarty:2002} where in addition to the $d$-wave
vertices in Fig.~\ref{fig:6vert} (a)-(b) one considers the effects of the $p$-wave vertices
in Fig.~\ref{fig:6vert} (c)-(f).~\cite{Nayak:2000} Such a description was shown to have a finite temperature phase transition without any specific heat anomaly. As it will become evident below,
one would have to include additional vertices to describe the transition to a
bond-incommensurate DDW state.

\subsection{Incommensurate}
\begin{figure}[htb]
\centerline{\includegraphics[height=3.0in]{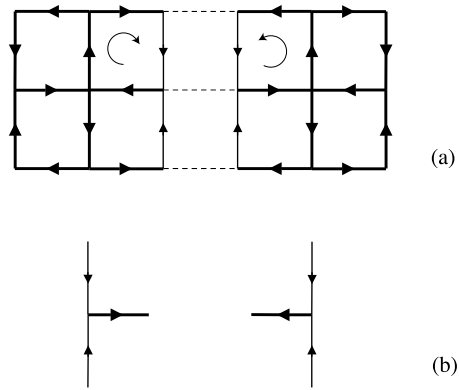}}
\caption{(a) A bond-oriented domain wall. (b) The additional vertices ensuring
current conservation at the domain boundary.} 
\label{fig:domainbond}
\end{figure}

To obtain an incommensurate version of DDW order, one first notes that since it breaks 
the Z$_2$ time-reversal symmetry, one can construct antiphase-domain walls 
of mutually reversed current configurations.
In Fig.~\ref{fig:domainbond} (a) we have an example of a bond-oriented domain 
wall. It can be written as,
\begin{equation}\label{domain_single}
\langle{c}^{\dagger}_{\bf R'}c_{\bf R}\rangle
= i(-1)^{m'+n'}\frac{W_0}{2}{\rm V}_{\bf R',R}\times\Theta(ma)
+i{\rm V}^{\rm edge}_{\bf R',R},
\end{equation}
where
\begin{equation}
\Theta(x) =\Biggl\{
\begin{matrix}
1 & , & x > 0 \\
0 & , & x = 0 \\
-1 & , & x < 0 
\end{matrix}
\end{equation}
is an overall antiphase modulation. The additional vertices 
${\rm V}^{\rm edge}_{\bf R',R}$ in Eq.~\eqref{domain_single} are necessary 
to ensure current conservation along the domain wall boundary where 
the current is {\it half} of that in the bulk. Explicitly, they are given by the following:
\begin{equation}\label{vert_ic}
\begin{split}
{\rm V}^{\rm edge}_{\bf R',R} &= {\rm V}^{\rm L}_{\bf R',R}\delta_{m,0}
+ {\rm V}^{\rm R}_{\bf R',R}\delta_{m,1}\\
{\rm V}^{\rm L}_{\bf R',R} & = \delta_{{\bf R',R}-\hat{\bf x}a} -
	\frac{1}{2}\delta_{{\bf R',R}+\hat{\bf y}a}-\frac{1}{2}\delta_{{\bf R',R}-\hat{\bf y}a}\\
{\rm V}^{\rm R}_{\bf R',R} & = \delta_{{\bf R',R}+\hat{\bf x}a} -
	\frac{1}{2}\delta_{{\bf R',R}+\hat{\bf y}a}-\frac{1}{2}\delta_{{\bf R',R}-\hat{\bf y}a}
\end{split}
\end{equation}
and where ${\rm V}^{\rm L}_{\bf R',R}$ and ${\rm V}^{\rm R}_{\bf R',R}$ are
shown in Fig. \ref{fig:domainbond} (b). 

\begin{figure}[htb]
\centerline{\includegraphics[height=3.0in]{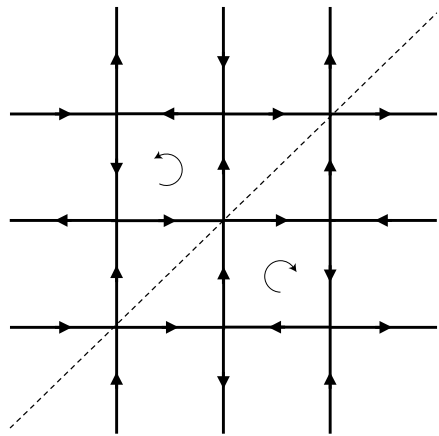}}
\caption{A diagonal domain wall. It can be entirely
constructed from the six vertices in Fig.\ref{fig:6vert}.} 
\label{fig:domaindiag}
\end{figure}

Note that the vertices in Eqs.~\eqref{vert_ic}
violate both $d$-wave and $p$-wave symmetries and hence a transition to bond IC-DDW
order will not be in the universality class of the 6-vertex model. On the other hand,
a diagonal domain wall, as in Fig.~\ref{fig:domaindiag}, does not require extra vertices,
and hence the transition to diagonal IC-DDW does belong to the 6-vertex model
universality class. 

Because the  contribution of ${\rm V}^{\rm edge}_{\bf R',R}$ 
is important only near a sharp bond-domain wall boundary, we drop it from now on. We will 
shortly see that at least in the single harmonic approximation, 
this is justified, for it leads to current conservation up to quadratic
order in the inverse domain wall spacing. Specifically, we can consider
an array of antiphase-domain walls at positions $\pm{d/2},\pm{3d/2},\dots$. 
Because the array is periodic, we can focus on the region $-d\leq ma\leq{d}$,
where due to the two domains at $\pm{d/2}$ the antiphase modulation $\Theta(ma)$
in \eqref{domain_single} is replaced by:
\begin{equation}\label{all-harmonics}
\theta(ma+\frac{d}{2})-\theta(-ma+\frac{d}{2})-1
=\sum_{l-{\rm odd}}A_{l}\cos(\frac{\pi lma}{d})
\end{equation}
with $A_l = 4/\pi{l}$.

Dropping the short range conributions  ${\rm V}^{\rm edge}_{\bf R',R}$ and 
keeping the first harmonic~\cite{Schulz:1989,Schulz:1990} $l=1$ in Eq.~\eqref{all-harmonics} amounts to 
considering an order parameter of the form:
\begin{equation}\label{harmonic1_x}
\langle{c}^{\dagger}_{\bf R'}c_{\bf R}\rangle
=  i(-1)^{m' + n'}\frac{W_0}{2}\cos({\bf q}\cdot{\bf R}){\rm V}_{\bf R',R}  
\end{equation}
where the incommensurability wavevector  
is proportional to the inverse domain wall seperation, so that for the bond-oriented
domain in Eq.~\eqref{domain_single}, ${\bf q} = (\pi/d, 0)$. 
Note that we regain the commensurate case of Eq.~\eqref{corder_x} as $q\rightarrow0$
since all domains are pushed to infinity in this limit.
With the expression in Eq.~\eqref{harmonic1_x} we can therefore analyze a continuous
transition to an IC state. It remains to check that the boundary 
contributions ${\rm V}^{\rm edge}_{\bf R',R}$ 
are unimportant, by verifying that current is conserved at each site 
to quadratic order in $\bf q$. 

The current along a bond 
${\bf R}$ and ${\bf R} +a\hat{\bf{s}}$, where $\hat{\bf{s}}= \hat{\bf{x}}$ or $\hat{\bf{s}}= \hat{\bf{y}}$,
arising from the IC-DDW state, Eq.~\eqref{harmonic1_x} , is given by
\begin{multline}\label{current_r}
J_{{\bf R},{\bf R} +a\hat{\bf{s}}}=- \frac{W_0}{2}{\rm V}_{{\bf R}_i,{\bf R}_i +a\hat{\bf{s}}}(-1)^{m+n}\times\\
\left[\cos{\bf q}\cdot({\bf R} +a\hat{\bf{s}})+\cos({\bf q}\cdot{\bf R})\right]. 
\end{multline}
 Therefore, the total current flowing out of a site vanishes up to quadratic order in $\bf q$.

In the  momentum space, the single-harmonic expression, Eq.~ \eqref{harmonic1_x} , looks particularly simple:
\begin{equation}\label{harmonic1_k}
\langle{c}^{\dagger}_{\bf k'}c_{{\bf k}}\rangle
= iW_{\bf k}\left[\delta_{{\bf k'},{\bf k+Q}}+\delta_{{\bf k'},{\bf k-Q}}\right].
\end{equation}
As the ordering wavevector ${\bf Q}=(\pi/a,\pi/a)+{\bf q}$ in 
Eq.~\eqref{harmonic1_k} becomes commensurate,
the above order reduces to that in Eq.~\eqref{corder_k}. Moreover, 
because DDW is odd with respect to rotations
by $\pi/2$ and transpositions about the $(\pi,\pi)$ direction one 
can check that in addition to domain walls,
a {\it checkerboard pattern} results from the following momentum-space representation:
\begin{equation}\label{checkerboard}
\langle{c}^{\dagger}_{\bf k'}c_{{\bf k}}\rangle
= \frac{i}{2}W_{\bf k}\left[\delta_{{\bf k'},{\bf k+Q}}+\delta_{{\bf k'},{\bf k-Q}}\right]
-\left[ {\bf Q}\rightarrow\mathcal{O}({\bf Q} )\right]
\end{equation}
where the operation $\mathcal{O}({\bf Q})$ can be either a transposition,
or a rotaion by $\pi/2$. Physically, the checkerboard pattern in 
Eq.~\eqref{checkerboard} corresponds to simply superimposing two
domain walls rotated by $\pi/2$ with respect to each other. 
In deriving Eq.~\eqref{harmonic1_k} we have considered
bond-oriented domains, but the construction is easily 
generalizable for the diagonal domain walls in 
Fig. \ref{fig:domaindiag}.

\section{Fermiology}
\subsection{Band structure}
We need to choose the band structure of the unreconstructed Fermi surface.  Although this
is not precisely known, it is generally accepted that it should contain
at least a nearest-neighbor, a next-nearest neighbor  and a third-neighbor matrix element in a tight binding Hamiltonian.~\cite{Andersen:1995} Thus, a sufficiently general form is
\begin{multline}\label{dispersion}
\epsilon_{\bf k}=-2t(\cos{k_x}a+\cos{k_y}a)+4t'\cos{k_x}a\cos{k_ya}\\
-2t''(\cos{2k_x}a+\cos{2k_y}a)
\end{multline}
In the past one of us  has chosen~\cite{Chakravarty:2008b} the parameters to be $t=0.3\; \text{eV}$, $t'=0.3 t$, and $t''=t'/9.0$,
 and to be consistent we shall adhere to these, although other reasonable choices will
 not change our main conclusions in the least; we could even set $t''=0$ or set it to a larger number, such as $t''=0.16 t$.~\cite{Millis:2007} This ambiguity is understandable because
 the precise results are not known, either empirically or theoretically, especially  in the underdoped regime.  A more serious problem is to correctly reconcile  the picture of the Fermi surface emerging from the  quantum oscillation observations with the angle resolved photoemission spectroscopy (ARPES).~\cite{Damascelli:2003} Finally, we will  have to choose a chemical potential, $\mu$, to attain the carrier concentration 
 we desire, and this we will mostly take to be about $10\%$ of holes, simply because many 
 quantum oscillation experiments 
 are available for this doping level. However, we will comment on what our theory yields as the 
 the doping increases or decreases.

\subsection{Strict incommensurability}
In this subsection we address the situation when strict incommensurability holds. For example,
${\bf Q} = {\bf Q}_{0}+{\bf q}=(\pi/a,\pi/a)-\pi({2\eta,0})/a$, where $\eta$ is not a rational number. It is of course a 
slight abuse of terminology to add the adjective `strict' but it dispels any possible sources of confusion. In this 
case each and every $\bf k$ and $\bf k+ Q$ correspond to a distinct point in the full Brillouin zone (BZ) of the underlying crystal lattice, and the fermion operators are similarly distinct. When necessary all sums will 
be carried out over the full BZ of the lattice with the conventional reciprocal lattice vectors.

Consider now the mean field DDW Hamiltoniaan given by
\begin{equation}
{\cal H} = \sum_{\bf k}\Psi^{\dagger}_{\bf k} X_{\bf k} \Psi_{\bf k}
\end{equation}
where
\begin{equation}
 X_{\bf k}\equiv
\begin{pmatrix}
\epsilon_{\bf k}-\mu & iG_{\bf k} & 0 & 0\\
\text{c.c.} & \epsilon_{\bf k+Q}-\mu & 0 & 0\\
0 & 0 & \epsilon_{\bf k}-\mu & -iG_{\bf k}^{'}\\
0 & 0 & \text{c.c.}& \epsilon_{\bf k+Q'}-\mu
\end{pmatrix}
\label{eq:ICDDW}
\end{equation}
and  
\begin{eqnarray}
G_{\bf k} &=& (W_{\bf k} - W_{\bf k+Q})/2\\
G_{\bf k}^{'} &=& (W_{\bf k} - W_{\bf k+Q^{'}})/2
\end{eqnarray}
The four component spinor is 
$\Psi_{\bf k}^{\dagger}\equiv(c^\dagger_{{\bf k}\uparrow},
c^\dagger_{{\bf k+Q}\uparrow},c_{{\bf k}\downarrow}^{\dagger},c^{\dagger}_{{\bf k+Q'}\downarrow})$.
It is useful to elaborate on this mean field Hamiltonian.
For a given wave vector ${\bf Q}=(\pi/a,\pi/a)+{\bf q}$, the inversion symmetry is broken, while
the pure imaginary character of the order parameter (and the compex Hermitian character of the Hamiltonian) implies broken time reversal symmetry. Note 
that the product of inversion and time reversal must be preserved as a physical requirement. Because the order parameter is 
a singlet in the spin space, this implies a very a simple matrix structure of the Hartree-Fock DDW
Hamiltonian. The inversion conjugate of this order parameter, ${\bf Q'}=(\pi/a,\pi/a)-{\bf q}$, must belong
to the down spin sector if $\bf Q$ belongs to the up spin sector.
To further clarify the form of this Hamiltonian, note that the use of the full BZ allows us to appropriately shift the
wave vectors to use identities like
\begin{eqnarray}
&&\sum_{\bf k}c^{\dagger}_{{\bf k},\alpha}c_{{\bf k+Q},\alpha}(iG_{\bf k})+ \text{h. c}.\\ \nonumber
&&=\sum_{{\bf k},{\bf k}'} c^{\dagger}_{{\bf k}\alpha}c_{{\bf k'}\alpha} (iW_{\bf k})\left[\delta_{{\bf k}',{\bf k+Q}}+\delta_{{\bf k}',{\bf k-Q}}\right] + \text{h. c.}
\end{eqnarray}
for any given spin direction $\alpha$.  

\begin{figure}[htb]
\begin{center}
\includegraphics[width=\linewidth]{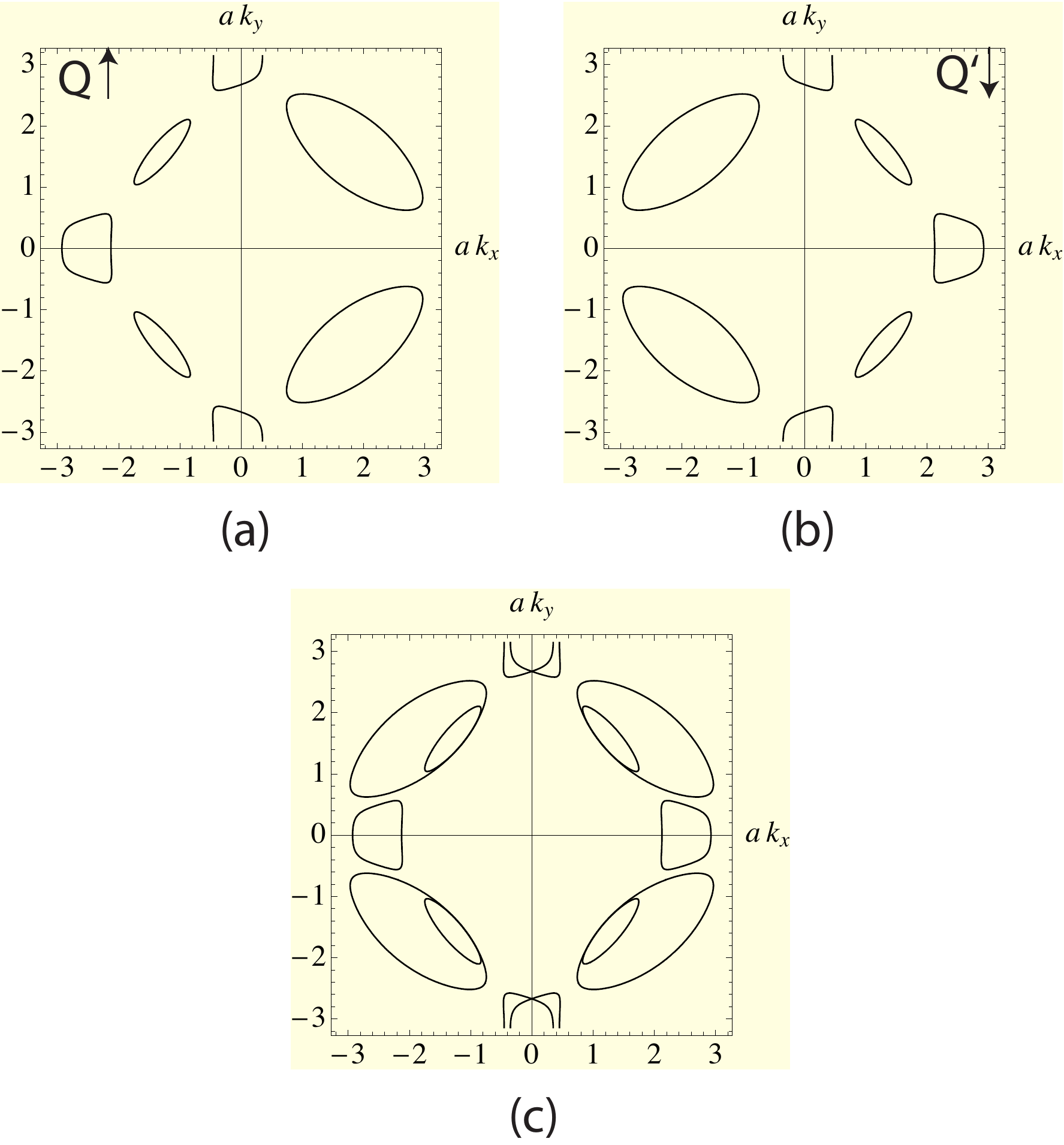}
\caption{(Color online) Reconstructed Fermi surfaces for incommensurate DDW order: (a) the up spin case with wave vector ${\bf Q}=(\pi/a,\pi/a)+{\bf q}$; (b) the down spin case with ${\bf Q}'=(\pi/a,\pi/a)-{\bf q}$; (c) the combined up and down spin spectra.}
\label{fig:IC-bands}
\end{center}
\end{figure}

The Fermi surfaces obtained from the spectra of Eq.~(\ref{eq:ICDDW}) are shown in Fig.~\ref{fig:IC-bands}. The pockets proximate to $(\pm\pi/a,\pm\pi/a)/2$ are hole pockets while those close to $(\pi/a,0)$ and its symmetry equivalents are electron pockets. To obtain this figure we have chosen: $W_{0}=0.112\; \text{eV}$,   $\eta=0.09$,~\cite{Dai:2001} $\mu=-0.256 \; \text{eV}$, and $n_{h}\approx 10\%$.
The corresponding dHvA frequencies determined by the Onsager relation 
\begin{equation}
F=\frac{\hbar c}{2\pi e}A(\epsilon_{F}), 
\end{equation}
where $A(\epsilon_{F})$ is the cross sectional area of a closed Fermi surface, are $F_{1}=526\; \text{T}$ (electron pocket), $F_{2}=1670\; \text{T}$ (large hole pocket), and $F_{3}=250\; \text{T}$ (small hole pocket). We have chosen the parameters to fit the data,~\cite{Sebastian:2008} but the existence and frequency of the small hole pocket is a prediction for which there appears to be very preliminary indications requiring further investigations and confirmation.~\cite{Sebastian:2008}

It is important to specify how we arrive at the total hole doping of $10\%$.  Fermi surface  is  a topological invariant.~\cite{Volovik:2003} Even when quasiparticles behave anomalously,   as in one-dimensional electronic systems, this surface is still defined by the same topological invariant. A break up of this surface, termed reconstruction,  requires a global deformation in the topological sense, most likely  a macroscopic broken symmetry.  The key here is Luttinger's sum rule. The most general form of this sum rule states that the particle density is twice (for two spin directions)  the volume of the wave vector  space in $d$-dimensions  divided by $(2\pi)^{d}$ over which the real part of the single particle Green function at the Fermi energy is positive, which applies even to Mott insulators.~\cite{Dzyaloshinskii:2003} This is easily applied here. There is one big hole pocket for a given spin direction within the {\em reconstructed} BZ. Considering the two spin directions producing identical pockets but reflected with respect to each other, the big hole pockets correspond to a fraction of carriers, $x_{\text{bh}}$, to be
\begin{equation}
x_{\text{bh}} = 2\frac{A_{\text{bh}}(\epsilon_{F})}{(2\pi/a)^{2}} ,
\label{eq:hdensity}
\end{equation}
in terms of carriers per Cu. Here $A_{\text{bh}}(\epsilon_{F})$ is the area of a big hole pocket. Note that the normalization in Eq. ~(\ref{eq:hdensity}) is with respect to the area of the {\em unreconstructed} BZ. Similarly, there is only one electron pocket in the {\em reconstructed} BZ and therefore the fraction of carriers, $x_{e}$, is, considering both spin partners,
\begin{equation}
x_{\text{e}} = 2\frac{A_{\text{e}}(\epsilon_{F})}{(2\pi/a)^{2}}.
\end{equation}  
But there is also one small hole pocket per {\em reconstructed} BZ and therefore
\begin{equation}
x_{\text{sh}} = 2\frac{A_{\text{sh}}(\epsilon_{F})}{(2\pi/a)^{2}}.
\label{eq:hdensity}
\end{equation}
The hole doping, $n_{h}$, is 
\begin{equation}
n_{h}=x_{\text{bh}}+x_{\text{sh}}-x_{\text{e}}.
\end{equation}
\begin{figure}[htb]
\begin{center}
\includegraphics[width=\linewidth]{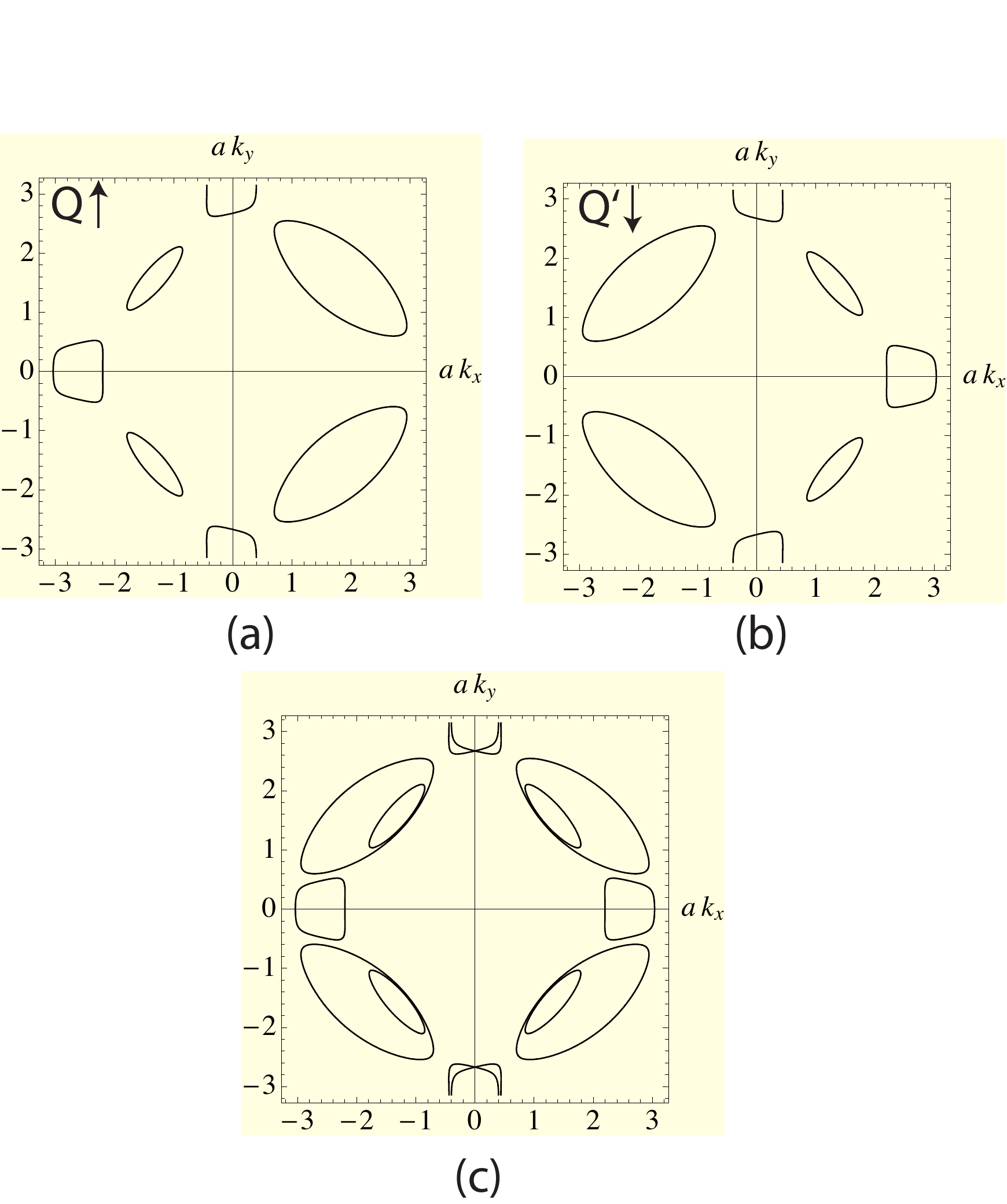}
\caption{(Color online) Reconstructed Fermi surafces for incommensurate spiral SDW order: (a) the up spin case with wave vector ${\bf Q}=(\pi/a,\pi/a)+{\bf q}$; (b) the down spin case with ${\bf Q}'=(\pi/a,\pi/a)-{\bf q}$; (c) the combined up and down spin spectra. Note that even though up and down spins are mixed we continue to label the spectra with the spin directions, since $\bf Q$ and $\bf Q'$ defines the cases unambiguously. }
\label{fig:IC-SDW-bands}
\end{center}
\end{figure}

It is important to compare the DDW case with the spiral SDW case. The mean field potential, $V_{\text{spiral}}$,  is
\begin{equation}
V_{\text{spiral}}=V_{0}\left[S_{x}\cos ({\bf Q}\cdot {\bf R})+S_{y}\sin ({\bf Q}\cdot {\bf R})\right].
\end{equation}
where $S_{x}$ and $S_{y}$ are the conventional spin half operators.
As pointed out by Overhauser,~\cite{Overhauser:1962,Daemen:1989} this interaction Hamiltonian breaks time reversal and inversion. It immediately follows that the mean field Hamiltonian matrix for the spiral SDW is
\begin{equation}
 Y_{\bf k}=
\begin{pmatrix}
\epsilon_{\bf k}-\mu & \Delta & 0 & 0\\
\Delta & \epsilon_{\bf k+Q}-\mu & 0 & 0\\
0 & 0 & \epsilon_{\bf k}-\mu & \Delta\\
0 & 0 & \Delta& \epsilon_{\bf k+Q'}-\mu
\end{pmatrix}
\label{eq:IC-SDW}
\end{equation}
where the four component spinor is
$\Phi_{\bf k}^{\dagger}\equiv(c^\dagger_{{\bf k}\uparrow},
c^\dagger_{{\bf k+Q}\downarrow},c_{{\bf k}\downarrow}^{\dagger},c^{\dagger}_{{\bf k+Q'}\uparrow})$. Here, $\Delta$, the spiral SDW gap parameter is of course real. But it is trivial to see that the eigenvalue structure is identical to the IC-DDW because both inversion and time reversal are broken. In contrast to IC-DDW, however,  the up and down sectors are mixed instead, as it should be, because DDW is a singlet in spin space, while SDW is a triplet, bearing in mind that both condensates are in the particle-hole channel. We can redefine, however, the  spinors, such as
\begin{equation}
\begin{pmatrix}
d_{{\bf k},\alpha}\\
d_{{\bf k+Q},\alpha}
\end{pmatrix}
=
\begin{pmatrix}
c_{{\bf k},\alpha}\\
i\epsilon_{\alpha\beta}c_{{\bf k+Q},\beta}
\end{pmatrix},
\end{equation}
similarly for ${\bf Q}\to {\bf Q}'$. The antisymmetric tensor $\epsilon_{\alpha\beta}$ is defined by $\epsilon_{11}\equiv\epsilon_{\uparrow\uparrow}=\epsilon_{22}\equiv\epsilon_{\downarrow\downarrow}=0$, $\epsilon_{12}\equiv\epsilon_{\uparrow\downarrow}=1$, and $\epsilon_{21}\equiv\epsilon_{\downarrow\uparrow}=-1$.
With the redefined spinors the  Hamiltonian matrix is now
\begin{equation}\label{matrix}
 Y_{\bf k}'=
\begin{pmatrix}
\epsilon_{\bf k}-\mu & i\Delta & 0 & 0\\
-i\Delta & \epsilon_{\bf k+Q}-\mu & 0 & 0\\
0 & 0 & \epsilon_{\bf k}-\mu &-i \Delta\\
0 & 0 & i\Delta& \epsilon_{\bf k+Q'}-\mu
\end{pmatrix},
\end{equation}
which closely resembles Eq.~(\ref{eq:ICDDW}).
The transformation clearly leaves the anticommutation rules unchanged. 

The Fermi surfaces resulting from Eq.~(\ref{eq:IC-SDW}) are shown in Fig.~\ref{fig:IC-SDW-bands}. They are essentially indistinguishable  from Fig.~\ref{fig:IC-bands}. This is despite the fact that the SDW order parameter is a triplet and is gapped, separating the valence band and the conduction band,  over the entire Fermi surface,   in contrast to the DDW order parameter. This is unfortunate because from dHvA frequencies, there is no way of telling if they arise from an incommensurate  spiral 
 SDW or an incommensurate DDW; one must invoke other considerations. In constructing this Fermi surface, we have chosen the following parameters: $\Delta=0.08\; \text{eV}$,   $\eta=0.08$, $\mu=-0.27 \; \text{eV}$, and $n_{h}\approx 10\%$. The corresponding dHvA frequencies are
  $F_{1}=533\; \text{T}$ (electron pocket), $F_{2}=1667\; \text{T}$ (large hole pocket), and $F_{3}=274\; \text{T}$ (small hole pocket). We suspect that a more diligent optimization of these parameters could be performed, but it will not change the main observations.

\subsection{Higher order commensurability}

As an example, consider higher order incommensuration of ${\bf  Q}={\bf Q}_{0}-\frac{\pi}{a}(2\times \frac{1}{8}, 0)=\frac{\pi}{a} (\frac{3}{4},1)$. 
\begin{widetext}
With the 8-component spinor defined by
$\chi_{\bf k}^{\dagger} = (c_{{\bf k}^{\dagger},\alpha }, c_{{\bf k+Q},\alpha }^{\dagger}, c_{{\bf k}+2{\bf Q},\alpha }^{\dagger}, \ldots c_{{\bf k}+8{\bf Q}}^{\dagger})$, the Hamiltonian can be written as 
\begin{equation}
{\cal H}=\sum_{{\bf k},\alpha} \chi_{{\bf k}\alpha}^{\dagger} Z_{{\bf k},\alpha}\chi_{{\bf k}\alpha}
\end {equation}
Now the up and down spin sector eigenvalues merely duplicate each other, and we can consider simply one of them:
\begin{equation}
Z_{\bf k}=\left(\begin{array}{cccccccc}\epsilon_{\bf k}-\mu & iG_{\bf k} & 0 & 0 & 0 & 0 & 0 & -iG_{{\bf k}+7{\bf Q}}  \\\text{c.c} & \epsilon_{\bf k+Q}-\mu & iG_{{\bf k}+{\bf Q}}  & 0 & 0 & 0 & 0 & 0\\0 & \text{c.c} & \epsilon_{{\bf k}+2{\bf Q}}-\mu & iG_{{\bf k}+2{\bf Q}} & 0 & 0 & 0 & 0 \\0 & 0 &  \text{c.c}  & \epsilon_{{\bf k}+3{\bf Q}}-\mu &  iG_{{\bf k}+3{\bf Q}}  & 0 & 0 & 0 \\0 & 0 & 0 & \text{c.c}  & \epsilon_{{\bf k}+4{\bf Q}}-\mu &  iG_{{\bf k}+4{\bf Q}}& 0 & 0 \\0 & 0 & 0 & 0 &  \text{c.c} & \epsilon_{{\bf k}+5{\bf Q}}-\mu & iG_{{\bf k}+5{\bf Q}} & 0 \\0 & 0 & 0 & 0 & 0 & \text{c.c} &  \epsilon_{{\bf k}+6{\bf Q}}-\mu &  iG_{{\bf k}+6{\bf Q}}  \\  iG_{{\bf k}+7{\bf Q}} & 0 & 0 & 0 & 0&0  &  \text{c.c} & \epsilon_{{\bf k}+7{\bf Q}}-\mu \end{array}\right).
\label{eq:highC}
\end{equation}
\end{widetext}

\begin{figure}[htb]
\begin{center}
\includegraphics[width=\linewidth]{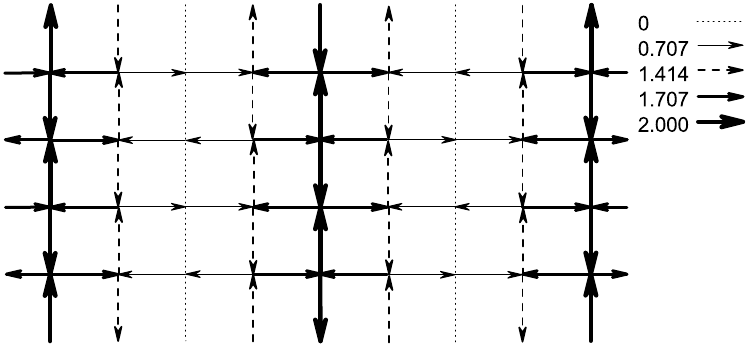}
\caption{Current pattern for ${\bf Q}=(\frac{3\pi}{ 4a}, \frac{\pi}{a})$. The relative magnitudes of the currents are depicted by the arrows in the legend.  Note the antiphase domain wall structure.}
\label{fig:IC-pattern}
\end{center}
\end{figure}

In the real space, we have
\begin{equation}
\label{cRR}
\begin{split}
    \langle c_{{\bf R}'}^\dag c_{\bf R}\rangle &= \frac{1}{N}\sum_{{\bf k}'{\bf k}}\langle c_{{\bf k}'}^\dag
   c_{{\bf k}}\rangle \exp\left[-i({\bf k}'\cdot{\bf R}'-{\bf k}\cdot{\bf R})\right]\\
   &=\pm \frac{i}{N}\sum_{{\bf k}}G_{\bf k}\exp\left[i{\bf k}\cdot({\bf R}-{\bf R}')\right]\exp\left[-i{\bf Q}\cdot{\bf R}'\right]\\
   &=\pm \frac{iW_0}{2}(-1)^{n'+m'}\left(\tilde{V}_{{\bf R}',{\bf R}}+i\tilde{U}_{{\bf R}',{\bf R}}\right),
\end{split}
\end{equation}
where ${\bf R}'=(m'a,n'a)$, and $\tilde{V}_{{\bf R}',{\bf R}}$ and $\tilde{U}_{{\bf
R}',{\bf R}}$ are of the form:\cite{Kee:2002}
\begin{equation}
\begin{split}
    \tilde{V}_{{\bf R}',{\bf R}}=&\Bigg[\frac{1+\cos 2\pi\eta}{2}(\delta_{{\bf R}',{\bf R}+a{\bf \hat{x}}}+\delta_{{\bf R}',{\bf R}-a{\bf\hat{x}}})\\
       & -(\delta_{{\bf R}',{\bf R}+a{\bf \hat{y}}}+\delta_{{\bf R}',{\bf R}-a{\bf\hat{y}}})\Bigg]\cos 2m'\pi\eta\\
        &+\frac{\sin 2\pi\eta\sin
        2m'\pi\eta}{2}(\delta_{{\bf R}',{\bf R}+a{\bf \hat{x}}}-\delta_{{\bf R}',{\bf R}-a{\bf\hat{x}}}),\\
    \tilde{U}_{{\bf R}',{\bf R}}=&-\Bigg[\frac{1+\cos 2\pi\eta}{2}(\delta_{{\bf R}',{\bf R}+a{\bf \hat{x}}}+\delta_{{\bf R}',{\bf R}-a{\bf\hat{x}}})\\
       & -(\delta_{{\bf R}',{\bf R}+a{\bf \hat{y}}}+\delta_{{\bf R}',{\bf R}-a{\bf\hat{y}}})\Bigg]\sin 2m'\pi\eta\\
        &+\frac{\sin 2\pi\eta\cos
        2m'\pi\eta}{2}(\delta_{{\bf R}',{\bf R}+a{\bf \hat{x}}}-\delta_{{\bf R}',{\bf R}-a{\bf\hat{x}}}).
\end{split}
\end{equation}
 We compute the current pattern from
\begin{equation}
\label{current}
\begin{split}
    J_{{\bf R}',{\bf R}}&=i[\langle c_{{\bf R}'}^\dag c_{\bf R}\rangle-\langle c_{\bf R}^\dag
        c_{{\bf R}'}\rangle]\\
    &=- W_{0}(-1)^{n'+m'}\tilde{V}_{{\bf R}',{\bf R}}
\end{split}
\end{equation}
As an example, the current pattern with $\eta=1/8$ is drawn in Fig.~\ref{fig:IC-pattern}. The relative
amplitudes of currents on bonds are labeled by different arrows.  Because $Q_y=\pi/a$, the
currents on a vertical slice have same amplitudes but alternating directions.
$Q_x=\pi/a-2\pi\eta/a=3\pi/4a$ results in a modulation of a period of $8$ lattice spacing along
the horizontal direction.

From  $\sum_{{\bf k}\in BZ}G_{\bf
k}=0$,  it follows that $\langle c_{\bf R}^\dag
c_{\bf R}\rangle =0$. This result is in disagreement with Ref.~\onlinecite{Kee:2002}, where the sum was incorrectly performed over the RBZ. Namely there is no site charge modulation in the higher order commensurate $d$-
density wave state. However, note that $\tilde{U}_{{\bf R}',{\bf R}}\ne 0$, for ${\bf R}\ne {\bf R}'$, and there are bond charge
modulations.

\begin{figure}[htb]
\begin{center}
\includegraphics[width=\linewidth]{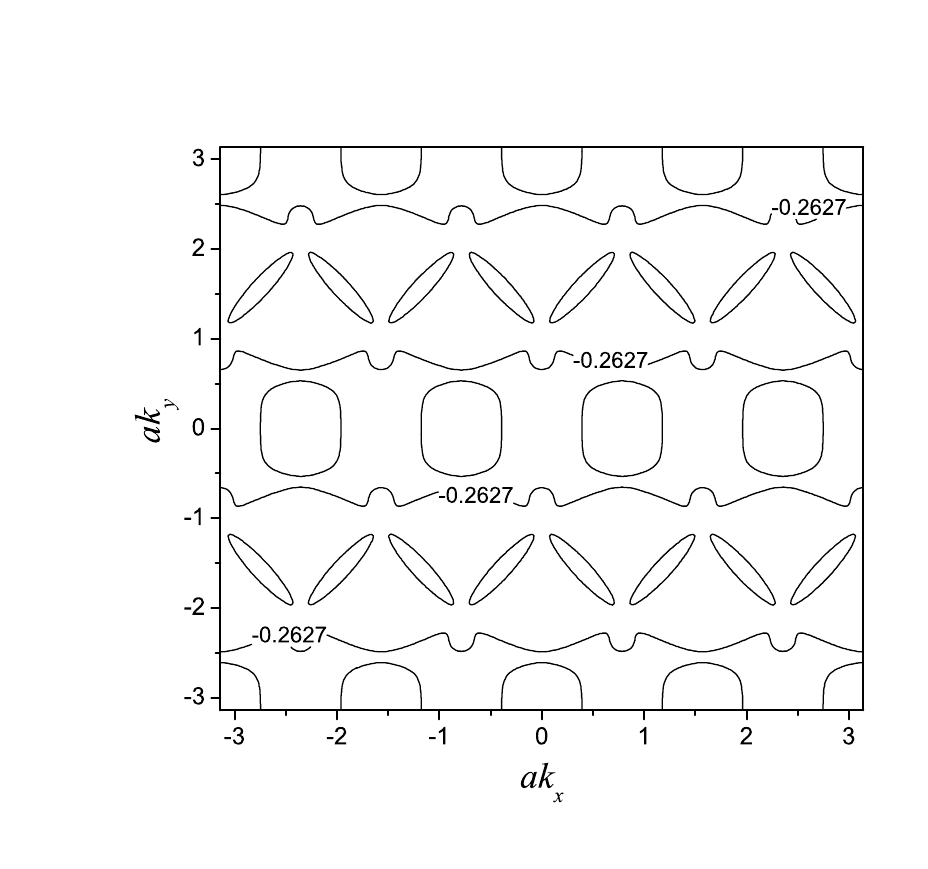}
\caption{Reconstructed Fermi surafces for a plausible high order commensurate DDW with ${\bf Q}=\frac{\pi}{a}(\frac{3}{4},1)$and $W_{0}=0.0825\; \text{eV}$. There are electron pockets, hole pockets and open orbits. Note that the figure is depicted in the extended BZ for clarity.}
\label{fig:highC}
\end{center}
\end{figure}

The Fermi surface corresponding to the spectra of Eq.~(\ref{eq:highC}) is shown in Fig.~\ref{fig:highC}. It is not essentially different from the  mean field theory of $1/8$ magnetic antiphase stripe order.~\cite{Millis:2007} This higher order commensuration  generically produces complicated Fermi surfaces, involving open orbits, hole pockets and electron pockets. However, it is difficult  to satisfy simultaneously the constraints of
the Luttinger sum rule,  the periodicity of the oscillations, and the negative sign of the Hall coefficient. It is not clear how this picture can be consistent with experiments, unless disorder, fluctuations, or magnetic breakdown~\cite{Falicov:1967} conspire delicately to reproduce the experimental observations, which of course can not be ruled out. We shall not pursue this approach further.

\subsection{Interlayer tunneling and bilayer splitting}
Bilayer coupling, $t_{\perp}({\bf k})$, in YBCO is  parametrized in terms of a momentum conserving tunneling matrix element. For tetragonal structure it is~\cite{Chakravarty:1993,Andersen:1995}
\begin{equation}
t_{\perp}({\bf k})=\frac{t_{\perp}}{4}\left[\cos (k_xa)-\cos (k_ya)\right]^2,
\label{eq:tmatrix1}
\end{equation}
where $a$ is the lattice spacing.
The tunneling Hamiltonian $H_{12}$ is given by
\begin{equation}
H_{12}=\sum_{{\bf k},\alpha}t_{\perp}({\bf k})(c_{{\bf
k},\alpha}^{\dagger (1)}c_{{\bf
k},\alpha}^{(2)}+1\leftrightarrow 2).
\end{equation}
The superscripts on the fermion operators refer to the layer index.
For simplicity consider first the commensurate case. The Hartree-Fock approximation, $H_{0}=H_{1}+H_{2}$ can be written
as the effective DDW Hamiltonians
\begin{equation}
\begin{split}
H_{0}&=\sum_{{\bf k} \in RBZ,\alpha}(\epsilon_{\bf k}c_{{\bf
k},\alpha}^{\dagger (1)}c_{{\bf
k},\alpha}^{(1)}+\epsilon_{{\bf k}+{\bf Q}_{0}}c_{{\bf
k}+\bf{Q}_{0},\alpha}^{\dagger (1)}c_{{\bf
k}+\bf{Q}_{0},\alpha}^{(1)}) \\ 
&+\sum_{{\bf k} \in RBZ,\alpha}(i \, W_{\mathbf{k}} c_{\mathbf{k} \alpha}^{\dagger (1)} c_{{\bf
k}+\bf{Q}_{0},\alpha}^{(1)}++\text{h. c.})\\
&+(1\leftrightarrow 2).
\end{split}
\label{eq:heff}
\end{equation} 
The reduced Brillouin zone (RBZ) is bounded by $k_{y}\pm k_{x}=\pm \pi/a$. With the choice of the quadratic Hamiltonian $H_{0}$ in Eq.~(\ref{eq:heff}), it can be easily diagonalized along with $H_{12}$. This is a first order  degenerate perturbation theory. Because $t_{\perp}$ will turn out to be so small, we do not expect a large correction.  

At each wave vector $k$ in the RBZ, we need to diagonalize a $4\times 4$ matrix to extract the energy eigenvalues. This matrix 
 is~\cite{Jia:2008}
\begin{equation}
\mathbb{H} = \left(\begin{array}{cccc}\epsilon_{\bf k} & iW_{\bf k} & t_{\perp} (\bf k)& 0 \\- iW_{\bf k} & \epsilon_{{\bf k}+{\bf Q}_{0}}  & 0 & t_{\perp} ({\bf k}+{\bf Q}_{0}) \\t_{\perp} (\bf k)& 0 & \epsilon_{\bf k} &  iW_{\bf k} \\0 & t_{\perp} ({\bf k}+{\bf Q}_{0}) & - iW_{\bf k}  & \epsilon_{{\bf k}+{\bf Q}_{0}} \end{array}\right).
 \label{eq:h}
\end{equation}
From $t_{\perp}({\bf k})$ it is clear that the electron pockets will be much more affected by it than the hole pockets. The bilayer splitting of the electron pocket  frequency in the dHvA meausurement, if it is to occur, should be smaller or of the order of the half-width at the half-maximum of the peak in the Fourier spectra, otherwise it would have been already resolved~\cite{Jaudet:2008,Sebastian:2008}. Combined with the Luttinger sum rule this provides a strongl constraint on the chosen parameters.

\begin{figure}
    \centering
  \includegraphics[width=7.5cm]{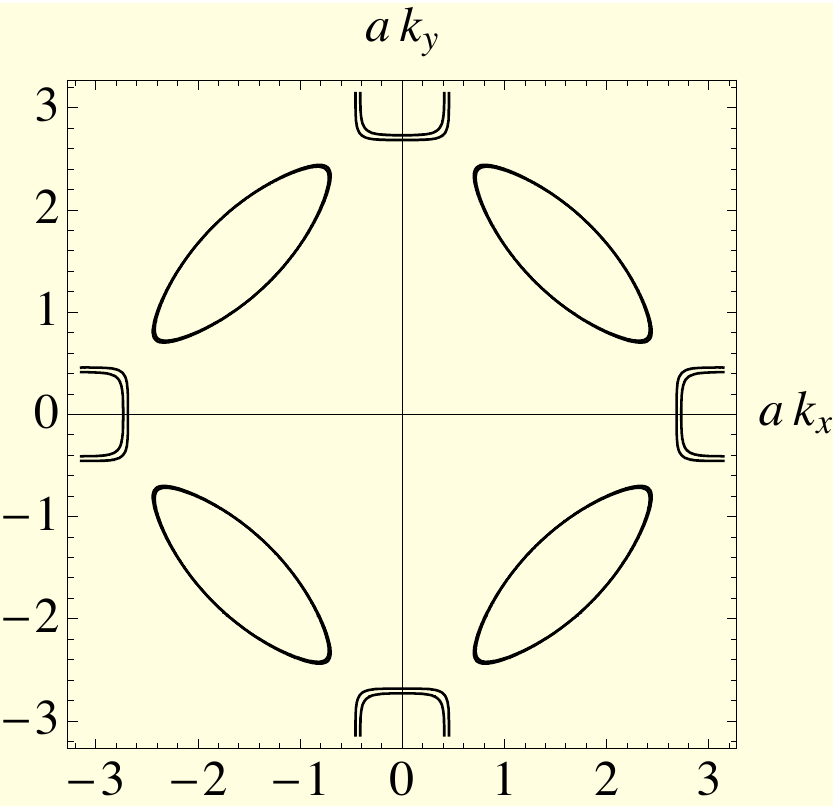}
  \caption{(Color online) Fermi surface split by bilayer coupling for commensurate DDW. The hole pockets centered at $\frac{1}{2}(\pm\pi/a,\pm\pi/a)$ are essentially unsplit by the bilayer coupling, In contrast, the electron pockets centered at  $(\pi/a,0)$ and symmetry related points are split as described in the text.}
  \label{fig:C}
\end{figure}

 As an illustration for  $\mathrm{YBa_{2}Cu_{3}O_{6.5}}$, we choose  $t_{\perp}=8\; \text{meV}$, and $W_{0}=0.0825\; \text{eV}$, and  the chemical potential $\mu$ is  set to $-0.2627\; \text{eV}$, which leads to a  hole doping of $n_{h}\approx 10\%$. The corresponding dHvA frequencies  are $F_{1}\approx 944\; \text{T}$, $F_{2}\approx 967\; \text{T}$, $F_{3}\approx 570\; \text{T}$, and $F_{4}\approx 450\; \text{T}$. The frequencies $F_{1}$ and $F_{2}$, corresponding  to the hole pockets are essentially unchanged  to our accuracy, while $F_{3}$ and $F_{4}$ correspond to the electron pockets split by the bilayer coupling. The Fermi surfaces are shown in Fig.~\ref{fig:C}.

The generalization to the incommensurate case is straightforward and has been discussed previously.~\cite{Jia:2008} Here we do not pursue this further but turn to a number of important conceptual issues. 
\begin{enumerate}
\item The interlayer tunneling Hamiltonian is real and cannot lead to orbital currents flowing between the layers. This is actually quite fortunate because the DDW order parameters within the planes remain undistorted. The sole effect of the interlayer Hamiltonian is to lead to linear superpositions of the independent quasiparticle states, which lead to a bilayer splitting, as in a generic two-state system.
\item If we are to choose opposite phasing of the orbital currents of the two layers, and change the matrix to  
\begin{equation}
\mathbb{H}' = \left(\begin{array}{cccc}\epsilon_{\bf k} & iW_{\bf k} & t_{\perp} (\bf k)& 0 \\- iW_{\bf k} & \epsilon_{{\bf k}+{\bf Q}_{0}}  & 0 & t_{\perp} ({\bf k}+{\bf Q}_{0}) \\t_{\perp} (\bf k)& 0 & \epsilon_{\bf k} &  -iW_{\bf k} \\0 & t_{\perp} ({\bf k}+{\bf Q}_{0}) &  iW_{\bf k}  & \epsilon_{{\bf k}+{\bf Q}_{0}} \end{array}\right),
 \label{eq:h}
\end{equation}
 there is no bilayer splitting  of the areas of the Fermi pockets. On the other hand, the ground state energy, to a very high degree of precision, is identical to the previous cases  even though the spectra are changed. We suspect that this equality is exact but have not been able to prove it analytically.
 Therefore, at the level of approximation discussed here, energetics do not allow  us  to choose between the alternatives   and the correct phasing should be determined from experiments.
 \item We have argued that the interlayer tunneling Hamiltonian is both real and bilinear and therefore does not result in a flow of current  between the layers. The only way such currents can appear is if there are further four fermion terms connecting the two layers such as particle-hole or particle-particle pair hopping terms. This would allow us to form further condensates made out of the two layers and would seriously distort the planar DDW order parameter. This has apparently gone unnoticed in Ref.~\onlinecite{Podolsky:2008}. The distortion of DDW order parameter must be thought through anew.
 \end{enumerate}

There is a further important consideration that is worth spelling out carefully. One of us recently argued~\cite{Kopp:2005} that the first order interalyer tunneling can be traded by a second order Hamiltonian in which virtual pair hopping of both particle-particle and particle-hole are present. The second order Hamiltonian involves four Fermi interactions from which appropriate condensates can be constructed. It was shown there that for such a Hamiltonian  antiferromagnetic arrangement of the orbital currents of the layers in a bilayer unit would be lower in energy than the ferromagnetic arrangement.  There were two arguments behind this effective Hamiltonian. The first consisted in noticing that the DDW quasiparticle spectra are substantially gapped at the antinodal points. Alas, given the existence of electron pockets in the quantum oscillation experiments, this mechanism could not be operative because there is no longer any gap at the antinodal points. The second argument involved non-Fermi liquid spectral function. One can easily see that with increased doping there would be Fermi surface reconnections near optimal doping and the electron pockets will evaporate. But at such doping, close to the middle of the superconducting dome, there will be a quantum critical point and one can expect the non-Fermi liquid behavior.~\cite{Daou:2008}  The systematics of the superconducting transition temperature of optimally doped homologous series can still be correctly given by the second order pair hopping terms.~\cite{Chakravarty:2004}

\section{Quantum oscillations in the mixed state}
There is considerable evidence that the quantum oscillation frequencies remain unshifted from the normal state in the mixed state of a wide class of superconductors, although there is increased damping arising from vortices.~\cite{Wasserman:1996} We shall closely follow Stephen~\cite{Stephen:1992} where this problem is solved for a conventional metal in the normal state and a $s$-wave superconductor with a gap $\Delta_{\text{s}}$. Since the Bogoliubov quasiparticles do not couple minimally to the gauge field, they do not form Landau levels and clearly can not be the source of quantum oscillations.~\cite{Franz:2000} The quantum oscillations must then arise from normal quasiparticles. By solving the Gorkov equations in the mixed state Stephen arrived at a formula for the self energy, which for high Landau levels, relevant for the present problem, is
\begin{equation}
\Sigma_{n}(i\omega)\approx \frac{ \Delta_{\text{s}}^2}{\sqrt{4\pi n}\hbar \omega_c}\left[-i\pi\textrm{sgn}(\omega)+\sqrt{\frac{\pi}{n}}\frac{\epsilon_n-\mu}{\hbar \omega_c}\right].
\label{NRLL18}
\end{equation}
where $\omega_{c}=\frac{eB}{m^{*}c}$ and $m^{*}$ is the corresponding effective mass.
The real part of the self-energy shifts the chemical potential and the Landau level positions equally. Hence, the real part of the self-energy does not affect the oscillation frequency. The imaginary part of the self-energy  leads to the following scattering rate
\begin{equation}
\frac{\hbar}{\tau_{v}}=\sqrt{\frac{\pi}{\nu}}\frac{\Delta_{\text{s}}^2}{\hbar \omega_c}=\Delta_{\text{s}}^2\sqrt{\frac{\pi}{|\mu| \hbar \omega_c}}= \Delta_{0}^2(1-\frac{B}{B_{c2}})\sqrt{\frac{\pi}{|\mu| \hbar \omega_c}},
\end{equation}
where the filling fraction is $\nu = |\mu|/(\hbar \omega_c)$ and $B_{c2}$ is the upper critical field.

To illustrate our main points, we shall treat the simpler commensurate case. The extension to the incommensurate case is straightforward. There are two cases to consider.  The effective mass corresponding to the electron pocket within the commensurate DDW theory, obtained by expanding around $(\pi/a,0)$, is  given by
\begin{equation}
\frac{\hbar^{2}}{2m^{*}}\approx \left(2t'+4t''-\frac{W_{0}}{4}\right) a^{2}.
\label{eq:emstar}
\end{equation}
Note that $t$ does not enter in the leading order. Assuming, $W_{0}=0.0825\; \text{eV}$, $\mu=-0.2642\; \text{eV}$, and $\Delta_{0}=10\; \text{meV}$,
we get 
\begin{equation}
\frac{1}{\tau_{v}}=2.9 \times 10^{12}\; \text{sec}^{-1},
\end{equation}
where we have used $B=40\; \text{T}$ and $B_{c2}=60\; \text{T}$ for the purpose of illustration. Using Eq.~(\ref{eq:emstar}) and the parameters given above, we find that $m^{*}=1.27 m_{e}$, where $m_{e}$ is the free electron mass.

A more interesting situation arises for the hole pocket, which, to an excellent approximation,  can be  described by the nodal fermions of DDW. For the purpose of illustration, we choose again the   commensurate case.  The Stephen formula has to be rederived for nodal fermions.

If the Fermi surface reconstruction is due to SDW, the excitations of both the electron and hole pockets can be approximated by the non-relativistic Schr\"odinger equations with the appropriate effective masses and Stephen's formula will apply without further modifications. As an aside, we note that for the spiral SDW given in Ref.~\onlinecite{Sebastian:2008}, the effective masses corresponding to both electron and hole pockets are considerably smaller (of the order of $0.5 \; m_{e}$) and do not seem to agree with the experimental results. Similarly, the Dingle factors for the spiral SDW are different from those of the  DDW order simply because SDW is gapped everywhere  and the excitations are given by non-relativistic fermions. 

\subsection*{Commensurate DDW}
The linearized two component Hamiltonian for DDW quasiparticles, corresponding to the node $(\pi/2a,\pi/2a)$, in the presence of an external magnetic field is given by
\begin{equation}
H=\int d\mathbf{r} \psi^{\dagger}(\mathbf{r})\hat{H}(\mathbf{r})\psi(\mathbf{r})
\end{equation}
where $\psi(\mathbf{r})$ is a two component spinor. The kernel $\hat{H}$ is given by
\begin{equation}
\hat{H}=v_F(p_x-eA_x/c)\sigma_3+v_D(p_y-eA_y/c)\sigma_2
\end{equation}
where $\sigma$'s are the standard Pauli matrices; $v_{F}$ is the velocity orthogonal to the Fermi surface and $v_{D}$ is the velocity tangential to it; $A_{x}$ and $A_{y}$ are the vector potentials and $c$ is the velocity of light.  The anisotropy between $x$ and $y$ directions can be removed with the redefinition of the coordinates $\tilde{x}=\sqrt{v_D/v_F}x$ and $\tilde{y}=\sqrt{v_F/v_D}y$. The measure of the integration remains unchanged, as $d\mathbf{r}= d\mathbf{\tilde{r}}$, and $\hat{H}$ becomes
\begin{equation}
\hat{H}=\sqrt{v_Fv_D}(p_{\tilde{x}}-eA_{\tilde{x}}/c)\sigma_3+\sqrt{v_Fv_D}(p_{\tilde{y}}-eA_{\tilde{y}}/c)\sigma_2
\end{equation}
Under the unitary transformation $\psi \to \mathrm{U}\Psi$, where $\mathrm{U}=(1+i\sigma_2)/\sqrt{2}$, $\hat{H}\to \hat{H}'$, such that 
\begin{multline}
\hat{H}^{\prime}=\sqrt{v_Fv_D}(p_{\tilde{x}}-eA_{\tilde{x}}/c)\sigma_1+\sqrt{v_Fv_D}(p_{\tilde{y}}-eA_{\tilde{y}}/c)\sigma_2
\end{multline}
Hence, the energy eigenvalues are 
\begin{equation}
\epsilon_{n,\tilde{k},\alpha}=\alpha \sqrt{n} \hbar \tilde{ \omega}=\alpha\hbar  \sqrt{2 n v_Fv_D}/l_{B}, 
\end{equation}
for $n\geq 1$, where $\alpha=\pm1$ stands for the particle and hole branches of the spectra. Here $\tilde{\omega}= \sqrt{2 v_Fv_D}/l_{B}$, $\tilde{k}=2\pi m/L_{\tilde{y}}$, $m=0,\pm1,\pm2,\ldots$, and the magnetic length $l_{B}=\sqrt{\hbar c/eB}$. The rescaled length $L_{\tilde{y}}=\sqrt{v_F/v_D}L_y$. For $n=0$, 
\begin{equation}
\epsilon_{0\tilde{k}}=\epsilon_{0}=0.
\end{equation}
\begin{widetext}
When $\textrm{sgn}(eB)>0$, the spinor wavefunctions in the Landau gauge $\mathbf{A}(\mathbf{\tilde{r}})=(0,B\tilde{x},0)$ are
\begin{eqnarray}
\Psi_{n\geq1,\tilde{k},\alpha}^{\dagger}(\mathbf{\tilde{r}})&=&\frac{1}{\sqrt{2}}(\varphi^{*}_{n,\tilde{k}}(\mathbf{\tilde{r}}),\alpha \varphi^{*}_{n-1,\tilde{k}}(\mathbf{\tilde{r}}))\\
\Psi_{n=0,\tilde{k}}^{\dagger}(\mathbf{\tilde{r}})&=&\frac{1}{\sqrt{2}}(\varphi^{*}_{0,{k}}(\mathbf{\tilde{r}}),0)
\end{eqnarray}
where
\begin{equation}
\varphi_{n,k}(\mathbf{\tilde{r}})=\left(\frac{1}{\sqrt{\pi}l_{B}L_{\tilde{y}}2^nn!}\right)^{\frac{1}{2}}\exp\left(-i\tilde{k}\tilde{y}-\frac{(\tilde{x}-\tilde{k}l_{B}^2)^2}{2l_{B}^2}\right)H_n\left(\frac{\tilde{x}-\tilde{k}l_{B}^2}{l_{B}}\right),
\end{equation}
and $H_{n}$ are the Hermite polynomials.
The matrix Green's function is defined as
\begin{equation}
G_{0}(\mathbf{\tilde{r}}_1,\mathbf{\tilde{r}}_2,i\omega)=\sum_{n\geq1,\tilde{k},\alpha}\frac{\Psi_{n,\tilde{k}}(\mathbf{\tilde{r}}_1)\Psi^{\dagger}_{n,\tilde{k}}(\mathbf{\tilde{r}}_2)}{i\omega-\epsilon_{n,\alpha}+\mu}+\sum_{\tilde{k}}\frac{\Psi_{0,\tilde{k}}(\mathbf{\tilde{r}}_1)\Psi^{\dagger}_{0,\tilde{k}}(\mathbf{\tilde{r}}_2)}{i\omega+\mu},
\end{equation}
After performing the sum over $\tilde{k}$ and $\alpha$, the Green's function can be expressed in terms of redefined coordinates as
\begin{eqnarray}
G_{0}(\mathbf{\tilde{r}}_1,\mathbf{\tilde{r}}_2,i\omega)=\frac{1}{2\pi l_{B}^2}\exp\left(-\frac{|\tilde{z}_{21}|^2}{4l_{B}^2}\right)\exp(i\phi_{21})\bigg[\sum_{n\geq1}\frac{i\omega+\mu}{(i\omega+\mu)^2-\epsilon_{n}^{2}}\left(\begin{array}{cc}
L_{n}\left(\frac{|\tilde{z}_{21}|^2}{2l_{B}^2}\right) & 0 \\
0 & L_{n-1}\left(\frac{|\tilde{z}_{21}|^2}{2l_{B}^2}\right) \end{array}\right)\nonumber \\
+\sum_{n\geq1}\frac{\epsilon_n}{l_{B}\sqrt{2n}[(i\omega+\mu)^2-\epsilon_{n}^{2}]}\left(\begin{array}{cc}
0 & -\tilde{z}_{21}L_{n-1}^{1}\left(\frac{|\tilde{z}_{21}|^2}{2l_{B}^2}\right) \\
\tilde{z}_{21}^{*}L_{n-1}^{1}\left(\frac{|\tilde{z}_{21}|^2}{2l_{B}^2}\right) & 0 \end{array}\right)+\frac{1}{i\omega+\mu}\left(\begin{array}{cc}
1 & 0 \\
0 & 0 \end{array}\right)\bigg]
\end{eqnarray}
where $\phi_{21}=(\tilde{x}_2+\tilde{x}_1)(\tilde{y}_2-\tilde{y}_1)/2l_{B}^2$, $\tilde{z}_{21}=(\tilde{x}_2-\tilde{x}_1)+i(\tilde{y}_2-\tilde{y}_1)$, and $L_{n}^{m}$ are the associate Laguerre polynomials. 
The Gorkov equation for the  normal component of the Green's function of a  $s$-wave superconductor is
\begin{equation}
G(\mathbf{\tilde{r}},\mathbf{\tilde{r}}^{\prime},i\omega)=G_0(\mathbf{\tilde{r}},\mathbf{\tilde{r}}^{\prime},i\omega)-\int d\mathbf{\tilde{r}}_1 d\mathbf{\tilde{r}}_2 G_0(\mathbf{\tilde{r}},\mathbf{\tilde{r}}_1,i\omega)\Delta_{\text{s}}(\mathbf{\tilde{r}}_1)G_0(\mathbf{\tilde{r}}_2,\mathbf{\tilde{r}}_1,-i\omega)\Delta_{\text{s}}^{*}(\mathbf{\tilde{r}}_2)G_0(\mathbf{\tilde{r}}_2,\mathbf{\tilde{r}}^{\prime},i\omega)
\end{equation}
For a disordered configuration of vortices (for other choices, see Stephen~\cite{Stephen:1992})
\begin{equation}
V(\mathbf{\tilde{r}}_1,\mathbf{\tilde{r}}_2)=\langle \Delta_{\text{s}}(\mathbf{\tilde{r}}_1)\Delta_{\text{s}}^{*}(\mathbf{\tilde{r}}_2) \exp(2i\phi_{21})\rangle=\Delta_{\text{s}}^{2}\exp\left(-\frac{(\mathbf{\tilde{r}}_2-\mathbf{\tilde{r}}_1)^2}{2l_{B}^2}\right).
\end{equation}
From the Gorkov equation we can identify the real space self-energy for the average Green's function to be
\begin{equation}
\Sigma(\mathbf{\tilde{r}}_1,\mathbf{\tilde{r}}_2)=V(\mathbf{\tilde{r}}_1,\mathbf{\tilde{r}}_2)\exp(2i\phi_{21})G_0(\mathbf{\tilde{r}}_2,\mathbf{\tilde{r}}_1,-i\omega).
\end{equation}
The matrix elements of $\Sigma(\mathbf{\tilde{r}}_1,\mathbf{\tilde{r}}_2)$ in the Landau level basis are given by
\begin{eqnarray}
\Sigma_{n_1,\tilde{k}_1,\alpha_1;n_2,\tilde{k}_2,\alpha_2}(i\omega)=\int d\mathbf{\tilde{r}}_1 d\mathbf{\tilde{r}}_2 \psi_{n_1,\tilde{k}_1,\alpha_1}^{\dagger}(\mathbf{\tilde{r}}_1)V(\mathbf{\tilde{r}}_1,\mathbf{\tilde{r}}_2)\exp(2i\phi_{21})G_0(\mathbf{\tilde{r}}_2,\mathbf{\tilde{r}}_1,-i\omega)\psi_{n_2,\tilde{k}_2,\alpha_2}(\mathbf{\tilde{r}}_2).
\end{eqnarray}
After performing the integrations we obtain
\begin{eqnarray}
\Sigma_{n_1,\tilde{k}_1,\alpha_1;n_2,\tilde{k}_2,\alpha_2}(i\omega)=\frac{\Delta_{\text{s}}^2}{2} \delta_{n_1,n_2}\delta_{\tilde{k}_1,\tilde{k}_2}\bigg[\sum_{n=0}^{\infty}\frac{(-i\omega+\mu)I_{n_1,n}}{(-i\omega+\mu)^2-\epsilon_{n}^{2}}+\alpha_1 \alpha_2\sum_{n=0}^{\infty}\frac{(-i\omega+\mu)I_{n_1-1,n}}{(-i\omega+\mu)^2-\epsilon_{n+1}^{2}}\nonumber \\-(\alpha_1+\alpha_2)\epsilon_{n_1}\sum_{n=0}^{\infty}\frac{I_{n_1,n}}{(-i\omega+\mu)^2-\epsilon_{n+1}^{2}}\bigg],
\end{eqnarray}
\end{widetext}
where
\begin{equation}
I_{n_1,n}=\frac{(n+n_1)!}{n!n_1!2^{n+n_1+1}}.
\label{NRLL16}
\end{equation}
When both $n$ and $n_1$ are large and comparable to the filling fraction $\nu \sim (|\mu|/\hbar \tilde{\omega})^{2}$,
\begin{equation}
I_{n_1,n}\sim \frac{1}{\sqrt{4\pi n}}\exp \left(-\frac{(n-n_1)^2}{4n}\right)
\label{NRLL17}
\end{equation}
and $\Sigma_{n}(i\omega)\sim \Sigma_{n_1}(i\omega)$.
For large $n$ and $n_1$, using the asymptotic form of $I_{n_1,n}$, and converting the the Landau level sums into integrals we find
\begin{multline}
\Sigma_{n,\alpha_1,\alpha_2}(i\omega)\approx\frac{\Delta_{\text{s}}^2 \mu\sqrt{\pi}}{4\sqrt{n}(\hbar \tilde{\omega})^2}\left\{\frac{1}{\sqrt{\pi n}}\frac{\mu^2-\epsilon_{n}^{2}}{(\hbar \tilde{\omega})^2}-i\ \textrm{sgn}(\omega \mu)
\right\}\\(1-\alpha_{1})(1-\alpha_{2}).
\end{multline}

It should be noted that for only the hole branch of the spectra is the self-energy nonzero. When $n \sim \nu$, we can neglect the real part of the self-energy and find the vortex contribution to the scattering rate to be
\begin{equation}
\frac{\hbar}{\tau_{v}}=\sqrt{\frac{\pi}{4}}\frac{\Delta_{\text{s}}^2}{\hbar \tilde{\omega}}=\sqrt{\frac{\pi}{8}}\frac{\Delta_{\text{s}}^2}{\sqrt{|\mu| \hbar \omega_{c}^{*}}}
\end{equation}
where the effective mass for DDW quasiparticles and the cyclotron frequency analogous to the non-relativistic case  are {\em defined dimensionally}  as 
\begin{eqnarray}
m^{*}&=&\frac{|\mu|}{v_Fv_D} ,
\label{eq:mstar}\\
\omega_{c}^{*}&=&\frac{eB}{m^{*}c}.
\end{eqnarray}
Note that being relativistic Weyl fermions, there is no real mass associated with them. Here $v_{F}=2\sqrt{2}at/\hbar$; to leading order neither $t'$ nor $t''$ enter. Similarly, $v_{D}=W_{0}a/\sqrt{2}\hbar$. Assuming, again,   $\Delta_{0}=10\;\text{meV}$, and $W_{0}=0.0825\; \text{eV}$,
we get
\begin{equation}
\frac{1}{\tau_{v}}=\frac{28}{\sqrt{B}}\left(1-\frac{B}{B_{c2}}\right)\times 10^{12} \text{sec}^{-1}
\end{equation}
where $B$ is in units of Tesla. For $B= 40\; \text{T}$ and $B_{c2}=60 \;\text{T}$, we get
\begin{equation}
\frac{1}{\tau_{v}}=1.5 \times 10^{12} \text{sec}^{-1}.
\end{equation}
Despite the relativistic nodal fermionic character of the quasiparticles, the quantum oscillation formulas at zero temperature, using the method of Ando,~\cite{Ando:1974} are formally identical to the non-relativistic case  with appropriate redefinitions. The damped oscillatory factor for the fundamental is~\cite{Goswami:2008}
\begin{equation}
e^{-\pi/\omega_{c}^{*}\tau_{v}} \cos\left(2\pi F/B\right),
\label{scba8}
\end{equation}
where $F$ is again given by the Onsager formula. If we also include the Dingle factor arising from the impurity scattering with a scattering time $\tau_{\text{i}}$, we need to multiply the above formula by a factor $e^{-\pi/\omega^{*}_{c}\tau_{\text{i}}}$. At a temperature $T$, the amplitude, ${\cal A}$,  is 
\begin{equation}
{\cal A}\propto \frac{\frac{2\pi^{2} k_{B}T}{\hbar\omega_{c}^{*}}}{\sinh (\frac{2\pi^{2} k_{B}T}{\hbar\omega_{c}^{*}})}
\end{equation}
We emphasize that $m^{*}$ here is a parameter, not the effective mass of the nodal fermions, which are actually massless. Thus, this formula could not be the conventional  Lifshitz-Kosevich formula derived for non-relativistic fermions, despite its formal similarity.  From Eq.~(\ref{eq:mstar}), the $m^{*}$ for the parameters given above is $2.72 \; m_{e}$, more than a factor of 2 larger than the non-relativistic mass corresponding to the carriers comprising the electron pocket. 

If we worked with a $d$-wave superconductor, then the Gorkov equation is more complicated due to the nonlocal gap function and there will be integration over four variables. However, the final answer for the scattering will not be significantly different from the $s$-wave answer. We can just replace $\Delta_{s}^2$ by the Fermi surface averaged $\langle|\Delta_{\bf k}|^{2}\rangle$, which is the same as averaging over extremal orbit in $2D$.

\section{Conclusions}\label{concl}
\subsection{Thoughts on microscopic models}
An effective  Hamiltonian  that can lead to both DDW and $d$-wave superconductivity (DSC)
in Hartree-Fock theory is the following:~\cite{Nayak:2002}
\begin{multline}
\mathcal{H}=-\sum_{ i,j}t_{ij}\left(
c_{i\alpha }^{\dagger }c_{j\alpha }^{}+\text{h.c.} \right)
- {t_c} {\hskip -0.3 cm}
\sum_{\substack{ \left\langle i,j\right\rangle ,
\left\langle i^{\prime},j\right\rangle \\i\neq i^{\prime} }}
c_{i\alpha }^{\dagger}c_{j\alpha }^{}c_{j\alpha' }^{\dagger }c_{i^{\prime}\alpha' }^{}\\
+ U\sum_in_{i\uparrow }^{}n_{j\downarrow}^{}
+V\sum_{\left\langle i,j\right\rangle }n_i^{}n_j^{}
\label{micro-Ham}
\end{multline}
In this formula, $t_{ij}$ is hopping matrix element with $t_{ij}=t$ for nearest
neighbors, $t_{ij}=-t^{\prime }$ for next nearest neighbors, etc. The operator $n_{i\alpha}$ is the density at a site $i$ and spin $\alpha$. On the other hand   $t_c$ is a correlated hopping term which simultaneously hops an electron
from site $j$ to site $i$ and hops an electron from $i'$ into the
vacated site $j$.
The on-site and nearest-neighbor repulsions are, respectively,
$U$ and  $V$. The repeated spin indices $\alpha, \alpha' $ are  assumed to be summed over.

At the mean field level,  the
energetics of the DDW order is described by the effective interaction
\begin{equation}\label{ddw-reduced}
\mathcal{H}_{DDW}=-g_\text{DDW}\int\limits_{\bf k,k'}f({\bf k})f({\bf k'})
c^\dagger_{{\bf k+Q}\alpha}c_{{\bf k}\alpha}
c^\dagger_{{\bf k'}\alpha'}c_{{\bf k'+Q}\alpha'},
\end{equation}
with $g_\text{DDW}=24t_c+8V$ and $f({\bf k})=(\cos k_{x}a - \cos k_{y} a)$.  A similar reduced Hamiltonian for DSC has the coupling  $g_\text{DSC}=12t_c-8V$.~\cite{Nayak:2002}
Note that the processes responsible for both DSC and DDW are essentially kinetic. 

A Hamiltonian given in Eq.~(\ref{micro-Ham})  has a $d$-wave
superconducting ground state over a range of dopings and  an antiferromagnetic ground
state at half-filling. Since this is the {\it sine qua non} for any description of the cuprates,
we believe that this is a good starting point for further calculations. Correlated hopping terms naturally arise even from the one-band Hubbard model away from half-filling,
where their contribution can be computed in $t/U$ perturbation theory: %
\begin{equation}
t_c \simeq n_{h}\frac{t^3}{U^2}
\end{equation}
At dopings $n_{h}=0.1$, for $t/U\sim0.2$, we would have
$g_\text{DDW}=2g_\text{DSC}\sim 0.1t\sim J/2$, $J$ being the antiferromagnetic exchange constant.
Thus, not far away from half-filling, correlated hopping terms are
certainly not negligible.

Preliminary calculations~\cite{Dimov:2008} at the Hartree-Fock level shows that it is difficult to make DDW incommensurate in the relevant underdoped  regime. The physical reason  is  that DDW has nodes. Our mean field results  suggest that
it costs less energy to dope holes by opening pockets around the nodes rather than
by making DDW incommensurate, as long as there is sufficient density of states left
at the antinodes to support commensurate DDW. 
To confirm the above intuition, we tested the effect of gapping the nodes by either
introducing $s$-wave order or including a $id_{xy}$ piece~\cite{Dimov:2008} to DDW.
 We could see that at finite
chemical potential, the effect of increasing the nodal gap created by the extra order
is to induce a first order transition to an incommensurate state when the nodal gap amplitude is
larger than a critical value, of the order of the $d_{x^2-y^2}$ amplitude itself. Of course Hartree-Fock 
approximation need not be the full answer because the notion of topological doping~\cite{Pryadko:1999} in 
the context of stripe order can be equally operative in the present context, a point that requires 
careful further work.

It is amusing to note that spiral SDW is also generically energetically unfavorable compared to collinear SDW in the weak coupling Hartree-Fock approximation, as  was pointed out by Overhauser.~\cite{Overhauser:1962} The strong coupling analysis can in principle lead to spirals, but these calculations are difficult to justify.~\cite{Shraiman:1989} 

\subsection{Experimental consequences}
If the assumption of incommensurate order is correct we would predict a number of interesting results. First, there should 
be four frequencies: the electron pocket should be split by the bilayer coupling, although this must be small because otherwise it would have been already resolved.  This sets the bilayer splitting in the range of $10-20 \; \text{meV}$ at the level of 10\% doping. In the overdoped regime the experimentally observed bilayer splitting of  $88 \; \text{meV}$  is already known to be substantially smaller than the estimate from the electronic structure calculations, which is about  $300 \; \text{meV}$.~\cite{Feng:2001} Why bilayer splitting is so small compared to electronic structure calculations is an important theoretical question requiring attention. To satisfy Luttinger sum rule, we not only predict a big hole pocket, but also a smaller hole pocket of frequency 250 T. Although there is a very preliminary indication,~\cite{Sebastian:2008} it needs to be investigated in further measurements and may constitute a confirming evidence of either a spiral SDW or an incommensurate DDW. 

IC-DDW could in principle be observed in Y-NMR measurements, as the fact that Y atom is situated at a high symmetry point is no longer an issue. However, the magnetic field at this site arising from the IC-DDW currents may be too small to observe this splitting. 

The most dramatic prediction of IC-DDW is the non-reflection symmetric band structure in a given spin sector.  A direct observation of such a band structure will be possible in a spin resolved ARPES measurement. Note that for spiral SDW up and down spins are mixed. Neutron scattering intensities from such quasiparticle bands may not be sufficient to detect it.

\subsection{Unresolved issues and future directions}

On the theoretical side a firm grasp on the existence of IC-DDW is necessary. This is beyond the scope of the present paper whose fundamental concern was to explore the phenomenological consequences of IC-DDW. As it stands, theory seems to favor the commensurate picture, although the predicted hole-pocket  in the commensurate case  is yet to be observed in quantum oscillation measurements. We expect that many of the experimental puzzles will be resolved in future experiments. It is of course desirable to go beyond Hartree-Fock to formulate a correct microscopic theory of IC-DDW. In particular, diagonal domain walls as a mechanism for incommensuration have not yet been explored. This seems necessary because it seems doubtful  that the antiphase domain walls that result in higher order commensurability, be it from DDW or antiphase spin stripe, can explain the present experiments.  Similarly, a rigorous theory of spiral SDW is desirable. As to IC-DDW, it is necessary to go beyond the approximation of  retaining only the single harmonic and to formulate a theory that fully conserves currents at the vertices.

The chief experimental  puzzle appears to be the conflict of the picture of the Fermi surface emerging from the quantum oscillations with the ARPES measurements.~\cite{Hossain:2008,Chang:2008} Although the hole pocket is not a major concern, because it has been argued that the back side of the hole pocket may not be visible because of the DDW coherence factors~\cite{Chakravarty:2003} and similar coherence factors in other forms of competing mechanism, the non-existence of the  electron pockets is more of a concern. From our previous calculations it is easy to see that at least two sides of the electron pockets should survive the effects of the coherence factors. It is noteworthy that in the past ARPES was able to detect both electron and hole pockets in electron doped cuprates.~\cite{Armitage:2001}

Although we do not believe that the observed negative Hall coefficient, its magnitude, and its oscillations can arise merely from the vortices in the mixed state, because it is difficult in such a scenario to obtain multiple Fermi pockets tightly constrained by the Luttinger sum rule, but it would be interesting to explore further the oscillations of the the Hall coefficient, in fact it  seems almost imperative. 

We have argued~\cite{Jia:2008} previously that the ortho-II potential of $\mathrm{YBa_{2}Cu_{3}O_{6.5}}$ is not an important factor in the explanation of the oscillations in contrast to the discussion in Ref.~\onlinecite{Podolsky:2008} . The highly polarizable BaO-layers next to the chains should screen the potentials quite effectively, to the extent that even disordered chains appear to have little effect in the planar physics in many properties. It is important to recognize that SdH, and dHvA,~\cite{Hussey:2008} measurements are also available in  $\mathrm{YBa_{2}Cu_{4}O_{8}}$ ($\sim 14\%$ doping~\cite{LeBoeuf:2007}), a stoichiometric compound with intrinsic oxygen order,  where even the negative Hall coefficient is clearly observed, in agreement with the observations in ortho-II materials.~\cite{LeBoeuf:2007} This degree of universality is unachievable if the chain potentials were playing an important role. 

A number of competing mechanisms~\cite{Lee:2008,Millis:2007,Chen:2008} to explain these unusual experiments have been proposed and have been discussed in a previous work by one of us.~\cite{Chakravarty:2008b} Thus, there is no need to duplicate the discussion here, but the salient questions that must be explained are worth repeating here. Are the oscillations, the negative sign, and the magnitude  of the Hall coefficient explainable without invoking a two-band scenario of a hole pocket and an electron pocket? Can the experimentally observed frequencies and the Luttinger sum rule be correctly reproduced? If the spiral spin density wave order or the antiphase IC-DDW order is the explanation what would be the defining experimental predictions that could be tested in further experiments? If the quasipartcles of a Fermi liquid are truly responsible for the oscillation measurements, what is the role of Mott physics in the cuprates? 

The lack of any definitive measurements vindicating  the static order parameters necessary to explain Fermi surface reconstruction (recall that dHvA is an equilibrium effect) is worth exploring. A magnetic field even as large as 60 T is not sufficient to provide enough perturbation to energetically nucleate static order in YBCO with an order parameter large enough to reproduce the measurements, especially in $\mathrm{YBa_{2}Cu_{4}O_{8}}$ with about $14\%$ doping. We hope that as other cuprates are explored and the theoretical tools are sharpened, many of these questions will be answered and will provide a resolution of the enigma of high-$T_{c}$.

\acknowledgements

We  thank  Reza Jamei and 
Eugene Pivovarov for helpful discussions.  We also acknowledge useful correspondence with 
our experimental colleagues: Louis Taillefer, Cyril Proust, S. Sebastian, Nigel Hussey, and Neil Harrison.
Special thanks are also due to Chetan Nayak, Bob Laughlin, Hae-Young Kee, Steve Kivelson and
Elihu Abrahams for a seemingly unending constructive dialogs. 
This work was supported by the NSF under Grant No.  DMR-0705092.


\begin{thebibliography}{63}
\expandafter\ifx\csname natexlab\endcsname\relax\def\natexlab#1{#1}\fi
\expandafter\ifx\csname bibnamefont\endcsname\relax
  \def\bibnamefont#1{#1}\fi
\expandafter\ifx\csname bibfnamefont\endcsname\relax
  \def\bibfnamefont#1{#1}\fi
\expandafter\ifx\csname citenamefont\endcsname\relax
  \def\citenamefont#1{#1}\fi
\expandafter\ifx\csname url\endcsname\relax
  \def\url#1{\texttt{#1}}\fi
\expandafter\ifx\csname urlprefix\endcsname\relax\def\urlprefix{URL }\fi
\providecommand{\bibinfo}[2]{#2}
\providecommand{\eprint}[2][]{\url{#2}}

\bibitem[{\citenamefont{Chakravarty et~al.}(1989)\citenamefont{Chakravarty,
  Halperin, and Nelson}}]{Chakravarty:1989}
\bibinfo{author}{\bibfnamefont{S.}~\bibnamefont{Chakravarty}},
  \bibinfo{author}{\bibfnamefont{B.~I.} \bibnamefont{Halperin}},
  \bibnamefont{and} \bibinfo{author}{\bibfnamefont{D.~R.}
  \bibnamefont{Nelson}}, \bibinfo{journal}{Phys. Rev. B}
  \textbf{\bibinfo{volume}{39}}, \bibinfo{pages}{2344} (\bibinfo{year}{1989}).

\bibitem[{\citenamefont{Kivelson et~al.}(2003)\citenamefont{Kivelson, Bindloss,
  Fradkin, Oganesyan, Tranquada, Kapitulnik, and Howald}}]{Kivelson:2003}
\bibinfo{author}{\bibfnamefont{S.~A.} \bibnamefont{Kivelson}},
  \bibinfo{author}{\bibfnamefont{I.~P.} \bibnamefont{Bindloss}},
  \bibinfo{author}{\bibfnamefont{E.}~\bibnamefont{Fradkin}},
  \bibinfo{author}{\bibfnamefont{V.}~\bibnamefont{Oganesyan}},
  \bibinfo{author}{\bibfnamefont{J.~M.} \bibnamefont{Tranquada}},
  \bibinfo{author}{\bibfnamefont{A.}~\bibnamefont{Kapitulnik}},
  \bibnamefont{and} \bibinfo{author}{\bibfnamefont{C.}~\bibnamefont{Howald}},
  \bibinfo{journal}{Reviews of Modern Physics} \textbf{\bibinfo{volume}{75}},
  \bibinfo{pages}{1201} (\bibinfo{year}{2003}).

\bibitem[{\citenamefont{Kivelson et~al.}(1998)\citenamefont{Kivelson, Fradkin,
  and Emery}}]{Kivelson:1998}
\bibinfo{author}{\bibfnamefont{S.~A.} \bibnamefont{Kivelson}},
  \bibinfo{author}{\bibfnamefont{E.}~\bibnamefont{Fradkin}}, \bibnamefont{and}
  \bibinfo{author}{\bibfnamefont{V.~J.} \bibnamefont{Emery}},
  \bibinfo{journal}{Nature} \textbf{\bibinfo{volume}{393}},
  \bibinfo{pages}{550} (\bibinfo{year}{1998}).

\bibitem[{\citenamefont{Varma}(1999)}]{Varma:1999}
\bibinfo{author}{\bibfnamefont{C.~M.} \bibnamefont{Varma}},
  \bibinfo{journal}{Phys. Rev. Lett.} \textbf{\bibinfo{volume}{83}},
  \bibinfo{pages}{3538} (\bibinfo{year}{1999}).

\bibitem[{\citenamefont{Chakravarty et~al.}(2001)\citenamefont{Chakravarty,
  Laughlin, Morr, and Nayak}}]{Chakravarty:2001}
\bibinfo{author}{\bibfnamefont{S.}~\bibnamefont{Chakravarty}},
  \bibinfo{author}{\bibfnamefont{R.~B.} \bibnamefont{Laughlin}},
  \bibinfo{author}{\bibfnamefont{D.~K.} \bibnamefont{Morr}}, \bibnamefont{and}
  \bibinfo{author}{\bibfnamefont{C.}~\bibnamefont{Nayak}},
  \bibinfo{journal}{Phys. Rev. B} \textbf{\bibinfo{volume}{63}},
  \bibinfo{pages}{094503} (\bibinfo{year}{2001}).

\bibitem[{\citenamefont{Caprara et~al.}(1999)\citenamefont{Caprara, Castellani,
  di~Castro, Grilli, and Sadori}}]{Caprara:1999}
\bibinfo{author}{\bibfnamefont{S.}~\bibnamefont{Caprara}},
  \bibinfo{author}{\bibfnamefont{C.}~\bibnamefont{Castellani}},
  \bibinfo{author}{\bibfnamefont{C.}~\bibnamefont{di~Castro}},
  \bibinfo{author}{\bibfnamefont{M.}~\bibnamefont{Grilli}}, \bibnamefont{and}
  \bibinfo{author}{\bibfnamefont{A.}~\bibnamefont{Sadori}}, in
  \emph{\bibinfo{booktitle}{Journal de Physique IV}} (\bibinfo{publisher}{EDP
  Sciences}, \bibinfo{year}{1999}), vol.~\bibinfo{volume}{9}, pp.
  \bibinfo{pages}{329--332}.

\bibitem[{\citenamefont{Zhang}(1997)}]{Zhang:1997}
\bibinfo{author}{\bibfnamefont{S.-C.} \bibnamefont{Zhang}},
  \bibinfo{journal}{Science} \textbf{\bibinfo{volume}{275}},
  \bibinfo{pages}{1089} (\bibinfo{year}{1997}).

\bibitem[{\citenamefont{Jaudet et~al.}(2008)\citenamefont{Jaudet, Vignolles,
  Audouard, Levallois, LeBoeuf, Doiron-Leyraud, Vignolle, Nardone, Zitouni,
  Liang et~al.}}]{Jaudet:2008}
\bibinfo{author}{\bibfnamefont{C.}~\bibnamefont{Jaudet}},
  \bibinfo{author}{\bibfnamefont{D.}~\bibnamefont{Vignolles}},
  \bibinfo{author}{\bibfnamefont{A.}~\bibnamefont{Audouard}},
  \bibinfo{author}{\bibfnamefont{J.}~\bibnamefont{Levallois}},
  \bibinfo{author}{\bibfnamefont{D.}~\bibnamefont{LeBoeuf}},
  \bibinfo{author}{\bibfnamefont{N.}~\bibnamefont{Doiron-Leyraud}},
  \bibinfo{author}{\bibfnamefont{B.}~\bibnamefont{Vignolle}},
  \bibinfo{author}{\bibfnamefont{M.}~\bibnamefont{Nardone}},
  \bibinfo{author}{\bibfnamefont{A.}~\bibnamefont{Zitouni}},
  \bibinfo{author}{\bibfnamefont{R.}~\bibnamefont{Liang}},
  \bibnamefont{et~al.}, \bibinfo{journal}{Physical Review Letters}
  \textbf{\bibinfo{volume}{100}}, \bibinfo{pages}{187005}
  (\bibinfo{year}{2008}).

\bibitem[{\citenamefont{Sebastian et~al.}(2008)\citenamefont{Sebastian,
  Harrison, Palm, Murphy, Mielke, Liang, Bonn, Hardy, and
  Lonzarich}}]{Sebastian:2008}
\bibinfo{author}{\bibfnamefont{S.~E.} \bibnamefont{Sebastian}},
  \bibinfo{author}{\bibfnamefont{N.}~\bibnamefont{Harrison}},
  \bibinfo{author}{\bibfnamefont{E.}~\bibnamefont{Palm}},
  \bibinfo{author}{\bibfnamefont{T.~P.} \bibnamefont{Murphy}},
  \bibinfo{author}{\bibfnamefont{C.~H.} \bibnamefont{Mielke}},
  \bibinfo{author}{\bibfnamefont{R.}~\bibnamefont{Liang}},
  \bibinfo{author}{\bibfnamefont{D.~A.} \bibnamefont{Bonn}},
  \bibinfo{author}{\bibfnamefont{W.~N.} \bibnamefont{Hardy}}, \bibnamefont{and}
  \bibinfo{author}{\bibfnamefont{G.~G.} \bibnamefont{Lonzarich}},
  \bibinfo{journal}{Nature} \textbf{\bibinfo{volume}{454}},
  \bibinfo{pages}{200} (\bibinfo{year}{2008}).

\bibitem[{\citenamefont{Yelland et~al.}(2008)\citenamefont{Yelland, Singleton,
  Mielke, Harrison, Balakirev, Dabrowski, and Cooper}}]{Yelland:2008}
\bibinfo{author}{\bibfnamefont{E.~A.} \bibnamefont{Yelland}},
  \bibinfo{author}{\bibfnamefont{J.}~\bibnamefont{Singleton}},
  \bibinfo{author}{\bibfnamefont{C.~H.} \bibnamefont{Mielke}},
  \bibinfo{author}{\bibfnamefont{N.}~\bibnamefont{Harrison}},
  \bibinfo{author}{\bibfnamefont{F.~F.} \bibnamefont{Balakirev}},
  \bibinfo{author}{\bibfnamefont{B.}~\bibnamefont{Dabrowski}},
  \bibnamefont{and} \bibinfo{author}{\bibfnamefont{J.~R.}
  \bibnamefont{Cooper}}, \bibinfo{journal}{Physical Review Letters}
  \textbf{\bibinfo{volume}{100}}, \bibinfo{pages}{047003}
  (\bibinfo{year}{2008}).

\bibitem[{\citenamefont{LeBoeuf et~al.}(2007)\citenamefont{LeBoeuf,
  Doiron-Leyraud, Levallois, Daou, Bonnemaison, Hussey, Balicas, Ramshaw,
  Liang, Bonn et~al.}}]{LeBoeuf:2007}
\bibinfo{author}{\bibfnamefont{D.}~\bibnamefont{LeBoeuf}},
  \bibinfo{author}{\bibfnamefont{N.}~\bibnamefont{Doiron-Leyraud}},
  \bibinfo{author}{\bibfnamefont{J.}~\bibnamefont{Levallois}},
  \bibinfo{author}{\bibfnamefont{R.}~\bibnamefont{Daou}},
  \bibinfo{author}{\bibfnamefont{J.~B.} \bibnamefont{Bonnemaison}},
  \bibinfo{author}{\bibfnamefont{N.~E.} \bibnamefont{Hussey}},
  \bibinfo{author}{\bibfnamefont{L.}~\bibnamefont{Balicas}},
  \bibinfo{author}{\bibfnamefont{B.~J.} \bibnamefont{Ramshaw}},
  \bibinfo{author}{\bibfnamefont{R.}~\bibnamefont{Liang}},
  \bibinfo{author}{\bibfnamefont{D.~A.} \bibnamefont{Bonn}},
  \bibnamefont{et~al.}, \bibinfo{journal}{Nature}
  \textbf{\bibinfo{volume}{450}}, \bibinfo{pages}{533} (\bibinfo{year}{2007}).

\bibitem[{\citenamefont{Doiron-Leyraud
  et~al.}(2007)\citenamefont{Doiron-Leyraud, Proust, LeBoeuf, Levallois,
  Bonnemaison, Liang, Bonn, Hardy, and Taillefer}}]{Doiron-Leyraud:2007}
\bibinfo{author}{\bibfnamefont{N.}~\bibnamefont{Doiron-Leyraud}},
  \bibinfo{author}{\bibfnamefont{C.}~\bibnamefont{Proust}},
  \bibinfo{author}{\bibfnamefont{D.}~\bibnamefont{LeBoeuf}},
  \bibinfo{author}{\bibfnamefont{J.}~\bibnamefont{Levallois}},
  \bibinfo{author}{\bibfnamefont{J.-B.} \bibnamefont{Bonnemaison}},
  \bibinfo{author}{\bibfnamefont{R.}~\bibnamefont{Liang}},
  \bibinfo{author}{\bibfnamefont{D.~A.} \bibnamefont{Bonn}},
  \bibinfo{author}{\bibfnamefont{W.~N.} \bibnamefont{Hardy}}, \bibnamefont{and}
  \bibinfo{author}{\bibfnamefont{L.}~\bibnamefont{Taillefer}},
  \bibinfo{journal}{Nature} \textbf{\bibinfo{volume}{447}},
  \bibinfo{pages}{565} (\bibinfo{year}{2007}).

\bibitem[{\citenamefont{Bangura et~al.}(2008)\citenamefont{Bangura, Fletcher,
  Carrington, Levallois, Nardone, Vignolle, Heard, Doiron-Leyraud, LeBoeuf,
  Taillefer et~al.}}]{Bangura:2008}
\bibinfo{author}{\bibfnamefont{A.~F.} \bibnamefont{Bangura}},
  \bibinfo{author}{\bibfnamefont{J.~D.} \bibnamefont{Fletcher}},
  \bibinfo{author}{\bibfnamefont{A.}~\bibnamefont{Carrington}},
  \bibinfo{author}{\bibfnamefont{J.}~\bibnamefont{Levallois}},
  \bibinfo{author}{\bibfnamefont{M.}~\bibnamefont{Nardone}},
  \bibinfo{author}{\bibfnamefont{B.}~\bibnamefont{Vignolle}},
  \bibinfo{author}{\bibfnamefont{P.~J.} \bibnamefont{Heard}},
  \bibinfo{author}{\bibfnamefont{N.}~\bibnamefont{Doiron-Leyraud}},
  \bibinfo{author}{\bibfnamefont{D.}~\bibnamefont{LeBoeuf}},
  \bibinfo{author}{\bibfnamefont{L.}~\bibnamefont{Taillefer}},
  \bibnamefont{et~al.}, \bibinfo{journal}{Physical Review Letters}
  \textbf{\bibinfo{volume}{100}}, \bibinfo{pages}{047004}
  (\bibinfo{year}{2008}).

\bibitem[{\citenamefont{Chakravarty}(2008)}]{Chakravarty:2008}
\bibinfo{author}{\bibfnamefont{S.}~\bibnamefont{Chakravarty}},
  \bibinfo{journal}{Science} \textbf{\bibinfo{volume}{319}},
  \bibinfo{pages}{735} (\bibinfo{year}{2008}).

\bibitem[{\citenamefont{Lee}(2008)}]{Lee:2008}
\bibinfo{author}{\bibfnamefont{P.~A.} \bibnamefont{Lee}},
  \bibinfo{journal}{Reports on Progress in Physics}
  \textbf{\bibinfo{volume}{71}}, \bibinfo{pages}{012501}
  (\bibinfo{year}{2008}).

\bibitem[{\citenamefont{Chakravarty and Kee}(2008)}]{Chakravarty:2008b}
\bibinfo{author}{\bibfnamefont{S.}~\bibnamefont{Chakravarty}} \bibnamefont{and}
  \bibinfo{author}{\bibfnamefont{H.-Y.} \bibnamefont{Kee}},
  \bibinfo{journal}{Proceedings of the National Academy of Sciences USA}
  \textbf{\bibinfo{volume}{105}}, \bibinfo{pages}{8835} (\bibinfo{year}{2008}).

\bibitem[{\citenamefont{Anderson}(1987)}]{Anderson:1987}
\bibinfo{author}{\bibfnamefont{P.~W.} \bibnamefont{Anderson}},
  \bibinfo{journal}{Science} \textbf{\bibinfo{volume}{235}},
  \bibinfo{pages}{1196} (\bibinfo{year}{1987}).

\bibitem[{\citenamefont{Halperin and Rice}(1968)}]{Halperin:1968}
\bibinfo{author}{\bibfnamefont{B.~I.} \bibnamefont{Halperin}} \bibnamefont{and}
  \bibinfo{author}{\bibfnamefont{T.~M.} \bibnamefont{Rice}}, in
  \emph{\bibinfo{booktitle}{Solid State Physics}}
  (\bibinfo{publisher}{Academic}, \bibinfo{address}{San Diego},
  \bibinfo{year}{1968}), p. \bibinfo{pages}{115}, \bibinfo{note}{edited by F.
  Seitz and D. Turnbull.}

\bibitem[{\citenamefont{Jia et~al.}(2008)\citenamefont{Jia, Dimov, Goswami, and
  Chakravarty}}]{Jia:2008}
\bibinfo{author}{\bibfnamefont{X.}~\bibnamefont{Jia}},
  \bibinfo{author}{\bibfnamefont{I.}~\bibnamefont{Dimov}},
  \bibinfo{author}{\bibfnamefont{P.}~\bibnamefont{Goswami}}, \bibnamefont{and}
  \bibinfo{author}{\bibfnamefont{S.}~\bibnamefont{Chakravarty}}
  (\bibinfo{year}{2008}), \urlprefix\url{http://arxiv.org/abs/0806.3793}.

\bibitem[{\citenamefont{Kopp et~al.}(2007)\citenamefont{Kopp, Ghosal, and
  Chakravarty}}]{Kopp:2007}
\bibinfo{author}{\bibfnamefont{A.}~\bibnamefont{Kopp}},
  \bibinfo{author}{\bibfnamefont{A.}~\bibnamefont{Ghosal}}, \bibnamefont{and}
  \bibinfo{author}{\bibfnamefont{S.}~\bibnamefont{Chakravarty}},
  \bibinfo{journal}{Proceedings of the National Academy of Sciences USA}
  \textbf{\bibinfo{volume}{104}}, \bibinfo{pages}{6123} (\bibinfo{year}{2007}).

\bibitem[{\citenamefont{Schollw\"ock et~al.}(2003)\citenamefont{Schollw\"ock,
  Chakravarty, Fj{\ae}restad, Marston, and Troyer}}]{Schollwoeck:2003}
\bibinfo{author}{\bibfnamefont{U.}~\bibnamefont{Schollw\"ock}},
  \bibinfo{author}{\bibfnamefont{S.}~\bibnamefont{Chakravarty}},
  \bibinfo{author}{\bibfnamefont{J.~O.} \bibnamefont{Fj{\ae}restad}},
  \bibinfo{author}{\bibfnamefont{J.~B.} \bibnamefont{Marston}},
  \bibnamefont{and} \bibinfo{author}{\bibfnamefont{M.}~\bibnamefont{Troyer}},
  \bibinfo{journal}{Physical Review Letters} \textbf{\bibinfo{volume}{90}},
  \bibinfo{pages}{186401} (\bibinfo{year}{2003}).

\bibitem[{\citenamefont{Emery et~al.}(1999)\citenamefont{Emery, Kivelson, and
  Tranquada}}]{Emery:1999}
\bibinfo{author}{\bibfnamefont{V.~J.} \bibnamefont{Emery}},
  \bibinfo{author}{\bibfnamefont{S.~A.} \bibnamefont{Kivelson}},
  \bibnamefont{and} \bibinfo{author}{\bibfnamefont{J.~M.}
  \bibnamefont{Tranquada}}, \bibinfo{journal}{Proceedings of the National
  Academy of Sciences USA} \textbf{\bibinfo{volume}{96}}, \bibinfo{pages}{8814}
  (\bibinfo{year}{1999}).

\bibitem[{\citenamefont{Stock et~al.}(2002)\citenamefont{Stock, Buyers, Tun,
  Liang, Peets, Bonn, Hardy, and Taillefer}}]{Stock:2002}
\bibinfo{author}{\bibfnamefont{C.}~\bibnamefont{Stock}},
  \bibinfo{author}{\bibfnamefont{W.~J.~L.} \bibnamefont{Buyers}},
  \bibinfo{author}{\bibfnamefont{Z.}~\bibnamefont{Tun}},
  \bibinfo{author}{\bibfnamefont{R.}~\bibnamefont{Liang}},
  \bibinfo{author}{\bibfnamefont{D.}~\bibnamefont{Peets}},
  \bibinfo{author}{\bibfnamefont{D.}~\bibnamefont{Bonn}},
  \bibinfo{author}{\bibfnamefont{W.~N.} \bibnamefont{Hardy}}, \bibnamefont{and}
  \bibinfo{author}{\bibfnamefont{L.}~\bibnamefont{Taillefer}},
  \bibinfo{journal}{Physical Review B} \textbf{\bibinfo{volume}{66}},
  \bibinfo{pages}{024505} (\bibinfo{year}{2002}).

\bibitem[{\citenamefont{Fauqu\'e et~al.}(2006)\citenamefont{Fauqu\'e, Sidis,
  Hinkov, Pailhes, Lin, Chaud, and Bourges}}]{Fauque:2006}
\bibinfo{author}{\bibfnamefont{B.}~\bibnamefont{Fauqu\'e}},
  \bibinfo{author}{\bibfnamefont{Y.}~\bibnamefont{Sidis}},
  \bibinfo{author}{\bibfnamefont{V.}~\bibnamefont{Hinkov}},
  \bibinfo{author}{\bibfnamefont{S.}~\bibnamefont{Pailhes}},
  \bibinfo{author}{\bibfnamefont{C.~T.} \bibnamefont{Lin}},
  \bibinfo{author}{\bibfnamefont{X.}~\bibnamefont{Chaud}}, \bibnamefont{and}
  \bibinfo{author}{\bibfnamefont{P.}~\bibnamefont{Bourges}},
  \bibinfo{journal}{Physical Review Letters} \textbf{\bibinfo{volume}{96}},
  \bibinfo{pages}{197001} (\bibinfo{year}{2006}).

\bibitem[{\citenamefont{Mook et~al.}(2002)\citenamefont{Mook, Dai, Hayden,
  Hiess, Lynn, Lee, and Do\v{g}an}}]{Mook:2002}
\bibinfo{author}{\bibfnamefont{H.~A.} \bibnamefont{Mook}},
  \bibinfo{author}{\bibfnamefont{P.}~\bibnamefont{Dai}},
  \bibinfo{author}{\bibfnamefont{S.~M.} \bibnamefont{Hayden}},
  \bibinfo{author}{\bibfnamefont{A.}~\bibnamefont{Hiess}},
  \bibinfo{author}{\bibfnamefont{J.~W.} \bibnamefont{Lynn}},
  \bibinfo{author}{\bibfnamefont{S.~H.} \bibnamefont{Lee}}, \bibnamefont{and}
  \bibinfo{author}{\bibfnamefont{F.}~\bibnamefont{Do\v{g}an}},
  \bibinfo{journal}{Physical Review B} \textbf{\bibinfo{volume}{66}},
  \bibinfo{pages}{144513} (\bibinfo{year}{2002}).

\bibitem[{\citenamefont{Mook et~al.}(2004)\citenamefont{Mook, Dai, Hayden,
  Hiess, Lee, and Do\v{g}an}}]{Mook:2004}
\bibinfo{author}{\bibfnamefont{H.~A.} \bibnamefont{Mook}},
  \bibinfo{author}{\bibfnamefont{P.}~\bibnamefont{Dai}},
  \bibinfo{author}{\bibfnamefont{S.~M.} \bibnamefont{Hayden}},
  \bibinfo{author}{\bibfnamefont{A.}~\bibnamefont{Hiess}},
  \bibinfo{author}{\bibfnamefont{S.~H.} \bibnamefont{Lee}}, \bibnamefont{and}
  \bibinfo{author}{\bibfnamefont{F.}~\bibnamefont{Do\v{g}an}},
  \bibinfo{journal}{Physical Review B} \textbf{\bibinfo{volume}{69}},
  \bibinfo{pages}{134509} (\bibinfo{year}{2004}).

\bibitem[{\citenamefont{Millis and Norman}(2007)}]{Millis:2007}
\bibinfo{author}{\bibfnamefont{A.~J.} \bibnamefont{Millis}} \bibnamefont{and}
  \bibinfo{author}{\bibfnamefont{M.~R.} \bibnamefont{Norman}},
  \bibinfo{journal}{Physical Review B} \textbf{\bibinfo{volume}{76}},
  \bibinfo{pages}{220503} (\bibinfo{year}{2007}).

\bibitem[{\citenamefont{Stephen}(1992)}]{Stephen:1992}
\bibinfo{author}{\bibfnamefont{M.~J.} \bibnamefont{Stephen}},
  \bibinfo{journal}{Physical Review B} \textbf{\bibinfo{volume}{45}},
  \bibinfo{pages}{5481} (\bibinfo{year}{1992}).

\bibitem[{\citenamefont{Nayak}(2000)}]{Nayak:2000}
\bibinfo{author}{\bibfnamefont{C.}~\bibnamefont{Nayak}},
  \bibinfo{journal}{Physical Review B} \textbf{\bibinfo{volume}{62}},
  \bibinfo{pages}{4880} (\bibinfo{year}{2000}).

\bibitem[{\citenamefont{Marston and Affleck}(1989)}]{Marston:1989}
\bibinfo{author}{\bibfnamefont{J.~B.} \bibnamefont{Marston}} \bibnamefont{and}
  \bibinfo{author}{\bibfnamefont{I.}~\bibnamefont{Affleck}},
  \bibinfo{journal}{Physical Review B} \textbf{\bibinfo{volume}{39}},
  \bibinfo{pages}{11538} (\bibinfo{year}{1989}).

\bibitem[{\citenamefont{Nersesyan et~al.}(1991)\citenamefont{Nersesyan,
  Japaridze, and Kimeridze}}]{Nersesyan:1991}
\bibinfo{author}{\bibfnamefont{A.~A.} \bibnamefont{Nersesyan}},
  \bibinfo{author}{\bibfnamefont{G.~I.} \bibnamefont{Japaridze}},
  \bibnamefont{and} \bibinfo{author}{\bibfnamefont{I.~G.}
  \bibnamefont{Kimeridze}}, \bibinfo{journal}{Journal of Physics-Condensed
  Matter} \textbf{\bibinfo{volume}{3}}, \bibinfo{pages}{3353}
  (\bibinfo{year}{1991}).

\bibitem[{\citenamefont{Chakravarty}(2002)}]{Chakravarty:2002}
\bibinfo{author}{\bibfnamefont{S.}~\bibnamefont{Chakravarty}},
  \bibinfo{journal}{Physical Review B} \textbf{\bibinfo{volume}{66}},
  \bibinfo{pages}{224505} (\bibinfo{year}{2002}).

\bibitem[{\citenamefont{Schulz}(1990)}]{Schulz:1990}
\bibinfo{author}{\bibfnamefont{H.~J.} \bibnamefont{Schulz}},
  \bibinfo{journal}{Physical Review Letters} \textbf{\bibinfo{volume}{64}},
  \bibinfo{pages}{1445} (\bibinfo{year}{1990}).

\bibitem[{\citenamefont{Schulz}(1989)}]{Schulz:1989}
\bibinfo{author}{\bibfnamefont{H.~J.} \bibnamefont{Schulz}},
  \bibinfo{journal}{Journal de Physique} \textbf{\bibinfo{volume}{50}},
  \bibinfo{pages}{2833} (\bibinfo{year}{1989}).

\bibitem[{\citenamefont{Andersen et~al.}(1995)\citenamefont{Andersen,
  Liechtenstein, Jepsen, and Paulsen}}]{Andersen:1995}
\bibinfo{author}{\bibfnamefont{O.~K.} \bibnamefont{Andersen}},
  \bibinfo{author}{\bibfnamefont{A.~I.} \bibnamefont{Liechtenstein}},
  \bibinfo{author}{\bibfnamefont{O.}~\bibnamefont{Jepsen}}, \bibnamefont{and}
  \bibinfo{author}{\bibfnamefont{F.}~\bibnamefont{Paulsen}},
  \bibinfo{journal}{Journal of Physics \& Chemistry of Solids}
  \textbf{\bibinfo{volume}{56}}, \bibinfo{pages}{1573} (\bibinfo{year}{1995}).

\bibitem[{\citenamefont{Damascelli et~al.}(2003)\citenamefont{Damascelli,
  Hussain, and Shen}}]{Damascelli:2003}
\bibinfo{author}{\bibfnamefont{A.}~\bibnamefont{Damascelli}},
  \bibinfo{author}{\bibfnamefont{Z.}~\bibnamefont{Hussain}}, \bibnamefont{and}
  \bibinfo{author}{\bibfnamefont{Z.-X.} \bibnamefont{Shen}},
  \bibinfo{journal}{Reviews of Modern Physics} \textbf{\bibinfo{volume}{75}},
  \bibinfo{pages}{473} (\bibinfo{year}{2003}).

\bibitem[{\citenamefont{Dai et~al.}(2001)\citenamefont{Dai, Mook, Hunt, and
  Do\v{g}an}}]{Dai:2001}
\bibinfo{author}{\bibfnamefont{P.}~\bibnamefont{Dai}},
  \bibinfo{author}{\bibfnamefont{H.~A.} \bibnamefont{Mook}},
  \bibinfo{author}{\bibfnamefont{R.~D.} \bibnamefont{Hunt}}, \bibnamefont{and}
  \bibinfo{author}{\bibfnamefont{F.}~\bibnamefont{Do\v{g}an}},
  \bibinfo{journal}{Physical Review B} \textbf{\bibinfo{volume}{63}},
  \bibinfo{pages}{054525} (\bibinfo{year}{2001}).

\bibitem[{\citenamefont{Volovik}(2003)}]{Volovik:2003}
\bibinfo{author}{\bibfnamefont{G.~E.} \bibnamefont{Volovik}},
  \emph{\bibinfo{title}{The universe in a Helium droplet}}
  (\bibinfo{publisher}{Cambridge University Press},
  \bibinfo{address}{Cambridge}, \bibinfo{year}{2003}).

\bibitem[{\citenamefont{Dzyaloshinskii}(2003)}]{Dzyaloshinskii:2003}
\bibinfo{author}{\bibfnamefont{I.}~\bibnamefont{Dzyaloshinskii}},
  \bibinfo{journal}{Physical Review B} \textbf{\bibinfo{volume}{68}},
  \bibinfo{pages}{085113} (\bibinfo{year}{2003}).

\bibitem[{\citenamefont{Overhauser}(1962)}]{Overhauser:1962}
\bibinfo{author}{\bibfnamefont{A.~W.} \bibnamefont{Overhauser}},
  \bibinfo{journal}{Physical Review} \textbf{\bibinfo{volume}{128}},
  \bibinfo{pages}{1437} (\bibinfo{year}{1962}).

\bibitem[{\citenamefont{Daemen and Overhauser}(1989)}]{Daemen:1989}
\bibinfo{author}{\bibfnamefont{L.~L.} \bibnamefont{Daemen}} \bibnamefont{and}
  \bibinfo{author}{\bibfnamefont{A.~W.} \bibnamefont{Overhauser}},
  \bibinfo{journal}{Physical Review B} \textbf{\bibinfo{volume}{39}},
  \bibinfo{pages}{6431} (\bibinfo{year}{1989}).

\bibitem[{\citenamefont{Kee and Kim}(2002)}]{Kee:2002}
\bibinfo{author}{\bibfnamefont{H.-Y.} \bibnamefont{Kee}} \bibnamefont{and}
  \bibinfo{author}{\bibfnamefont{Y.~B.} \bibnamefont{Kim}},
  \bibinfo{journal}{Physical Review B} \textbf{\bibinfo{volume}{66}},
  \bibinfo{pages}{012505} (\bibinfo{year}{2002}).

\bibitem[{\citenamefont{Falicov and Zuckermann}(1967)}]{Falicov:1967}
\bibinfo{author}{\bibfnamefont{L.~M.} \bibnamefont{Falicov}} \bibnamefont{and}
  \bibinfo{author}{\bibfnamefont{M.~J.} \bibnamefont{Zuckermann}},
  \bibinfo{journal}{Physical Review} \textbf{\bibinfo{volume}{160}},
  \bibinfo{pages}{372} (\bibinfo{year}{1967}).

\bibitem[{\citenamefont{Chakravarty et~al.}(1993)\citenamefont{Chakravarty,
  Sudb\o, Anderson, and Strong}}]{Chakravarty:1993}
\bibinfo{author}{\bibfnamefont{S.}~\bibnamefont{Chakravarty}},
  \bibinfo{author}{\bibfnamefont{A.}~\bibnamefont{Sudb\o}},
  \bibinfo{author}{\bibfnamefont{P.~W.} \bibnamefont{Anderson}},
  \bibnamefont{and} \bibinfo{author}{\bibfnamefont{S.}~\bibnamefont{Strong}},
  \bibinfo{journal}{Science} \textbf{\bibinfo{volume}{261}},
  \bibinfo{pages}{337} (\bibinfo{year}{1993}).

\bibitem[{\citenamefont{Podolsky and Kee}(2008)}]{Podolsky:2008}
\bibinfo{author}{\bibfnamefont{D.}~\bibnamefont{Podolsky}} \bibnamefont{and}
  \bibinfo{author}{\bibfnamefont{H.-Y.} \bibnamefont{Kee}}
  (\bibinfo{year}{2008}), \urlprefix\url{http://arxiv.org/abs/0806.0005}.

\bibitem[{\citenamefont{Kopp and Chakravarty}(2005)}]{Kopp:2005}
\bibinfo{author}{\bibfnamefont{A.}~\bibnamefont{Kopp}} \bibnamefont{and}
  \bibinfo{author}{\bibfnamefont{S.}~\bibnamefont{Chakravarty}}, in
  \emph{\bibinfo{booktitle}{Strongly Correlated Electron Materials: Physics and
  Nanoengineering}} (\bibinfo{publisher}{SPIE}, \bibinfo{address}{San Diego,
  CA, USA}, \bibinfo{year}{2005}), vol. \bibinfo{volume}{5932}, pp.
  \bibinfo{pages}{593219--13},
  \urlprefix\url{http://arxiv.org/abs/cond-mat/0507574}.

\bibitem[{\citenamefont{Daou et~al.}(2008)\citenamefont{Daou, LeBoeuf,
  Doiron-Leyraud, Li, Laliberte, Cyr-Choiniere, Jo, Balicas, Yan, Zhou
  et~al.}}]{Daou:2008}
\bibinfo{author}{\bibfnamefont{R.}~\bibnamefont{Daou}},
  \bibinfo{author}{\bibfnamefont{D.}~\bibnamefont{LeBoeuf}},
  \bibinfo{author}{\bibfnamefont{N.}~\bibnamefont{Doiron-Leyraud}},
  \bibinfo{author}{\bibfnamefont{S.~Y.} \bibnamefont{Li}},
  \bibinfo{author}{\bibfnamefont{F.}~\bibnamefont{Laliberte}},
  \bibinfo{author}{\bibfnamefont{O.}~\bibnamefont{Cyr-Choiniere}},
  \bibinfo{author}{\bibfnamefont{Y.~J.} \bibnamefont{Jo}},
  \bibinfo{author}{\bibfnamefont{L.}~\bibnamefont{Balicas}},
  \bibinfo{author}{\bibfnamefont{J.~Q.} \bibnamefont{Yan}},
  \bibinfo{author}{\bibfnamefont{J.~S.} \bibnamefont{Zhou}},
  \bibnamefont{et~al.} (\bibinfo{year}{2008}),
  \urlprefix\url{http://arxiv.org/abs/0806.4621}.

\bibitem[{\citenamefont{Chakravarty et~al.}(2004)\citenamefont{Chakravarty,
  Kee, and V\"olker}}]{Chakravarty:2004}
\bibinfo{author}{\bibfnamefont{S.}~\bibnamefont{Chakravarty}},
  \bibinfo{author}{\bibfnamefont{H.~Y.} \bibnamefont{Kee}}, \bibnamefont{and}
  \bibinfo{author}{\bibfnamefont{K.}~\bibnamefont{V\"olker}},
  \bibinfo{journal}{Nature} \textbf{\bibinfo{volume}{428}}, \bibinfo{pages}{53}
  (\bibinfo{year}{2004}).

\bibitem[{\citenamefont{Wasserman and Springford}(1996)}]{Wasserman:1996}
\bibinfo{author}{\bibfnamefont{A.}~\bibnamefont{Wasserman}} \bibnamefont{and}
  \bibinfo{author}{\bibfnamefont{M.}~\bibnamefont{Springford}},
  \bibinfo{journal}{Advances in Physics} \textbf{\bibinfo{volume}{45}},
  \bibinfo{pages}{471} (\bibinfo{year}{1996}).

\bibitem[{\citenamefont{Franz and Te\v{s}anovi\'c}(2000)}]{Franz:2000}
\bibinfo{author}{\bibfnamefont{M.}~\bibnamefont{Franz}} \bibnamefont{and}
  \bibinfo{author}{\bibfnamefont{Z.}~\bibnamefont{Te\v{s}anovi\'c}},
  \bibinfo{journal}{Physical Review Letters} \textbf{\bibinfo{volume}{84}},
  \bibinfo{pages}{554} (\bibinfo{year}{2000}).

\bibitem[{\citenamefont{Ando}(1974)}]{Ando:1974}
\bibinfo{author}{\bibfnamefont{T.}~\bibnamefont{Ando}},
  \bibinfo{journal}{Journal of the Physical Society of Japan}
  \textbf{\bibinfo{volume}{37}}, \bibinfo{pages}{1233} (\bibinfo{year}{1974}).

\bibitem[{\citenamefont{Goswami and Chakravarty}(2008)}]{Goswami:2008}
\bibinfo{author}{\bibfnamefont{P.}~\bibnamefont{Goswami}} \bibnamefont{and}
  \bibinfo{author}{\bibfnamefont{S.}~\bibnamefont{Chakravarty}}
  (\bibinfo{year}{2008}), \bibinfo{note}{unpublished.}

\bibitem[{\citenamefont{Nayak and Pivovarov}(2002)}]{Nayak:2002}
\bibinfo{author}{\bibfnamefont{C.}~\bibnamefont{Nayak}} \bibnamefont{and}
  \bibinfo{author}{\bibfnamefont{E.}~\bibnamefont{Pivovarov}},
  \bibinfo{journal}{Physical Review B} \textbf{\bibinfo{volume}{66}},
  \bibinfo{pages}{064508} (\bibinfo{year}{2002}).

\bibitem[{\citenamefont{Dimov}(2008)}]{Dimov:2008}
\bibinfo{author}{\bibfnamefont{I.}~\bibnamefont{Dimov}}, Ph.D. thesis,
  \bibinfo{school}{University of California Los Angeles}
  (\bibinfo{year}{2008}), \bibinfo{note}{this work supersedes
  http://arxiv.org/abs/cond-mat/0512627 that contained numerical and other
  errors.}

\bibitem[{\citenamefont{Pryadko et~al.}(1999)\citenamefont{Pryadko, Kivelson,
  Emery, Bazaliy, and Demler}}]{Pryadko:1999}
\bibinfo{author}{\bibfnamefont{L.~P.} \bibnamefont{Pryadko}},
  \bibinfo{author}{\bibfnamefont{S.~A.} \bibnamefont{Kivelson}},
  \bibinfo{author}{\bibfnamefont{V.~J.} \bibnamefont{Emery}},
  \bibinfo{author}{\bibfnamefont{Y.~B.} \bibnamefont{Bazaliy}},
  \bibnamefont{and} \bibinfo{author}{\bibfnamefont{E.~A.}
  \bibnamefont{Demler}}, \bibinfo{journal}{Physical Review B}
  \textbf{\bibinfo{volume}{60}}, \bibinfo{pages}{7541} (\bibinfo{year}{1999}).

\bibitem[{\citenamefont{Shraiman and Siggia}(1989)}]{Shraiman:1989}
\bibinfo{author}{\bibfnamefont{B.~I.} \bibnamefont{Shraiman}} \bibnamefont{and}
  \bibinfo{author}{\bibfnamefont{E.~D.} \bibnamefont{Siggia}},
  \bibinfo{journal}{Physical Review Letters} \textbf{\bibinfo{volume}{62}},
  \bibinfo{pages}{1564} (\bibinfo{year}{1989}).

\bibitem[{\citenamefont{Feng et~al.}(2001)\citenamefont{Feng, Armitage, Lu,
  Damascelli, Hu, Bogdanov, Lanzara, Ronning, Shen, Eisaki et~al.}}]{Feng:2001}
\bibinfo{author}{\bibfnamefont{D.~L.} \bibnamefont{Feng}},
  \bibinfo{author}{\bibfnamefont{N.~P.} \bibnamefont{Armitage}},
  \bibinfo{author}{\bibfnamefont{D.~H.} \bibnamefont{Lu}},
  \bibinfo{author}{\bibfnamefont{A.}~\bibnamefont{Damascelli}},
  \bibinfo{author}{\bibfnamefont{J.~P.} \bibnamefont{Hu}},
  \bibinfo{author}{\bibfnamefont{P.}~\bibnamefont{Bogdanov}},
  \bibinfo{author}{\bibfnamefont{A.}~\bibnamefont{Lanzara}},
  \bibinfo{author}{\bibfnamefont{F.}~\bibnamefont{Ronning}},
  \bibinfo{author}{\bibfnamefont{K.~M.} \bibnamefont{Shen}},
  \bibinfo{author}{\bibfnamefont{H.}~\bibnamefont{Eisaki}},
  \bibnamefont{et~al.}, \bibinfo{journal}{Physical Review Letters}
  \textbf{\bibinfo{volume}{86}}, \bibinfo{pages}{5550} (\bibinfo{year}{2001}).

\bibitem[{\citenamefont{Hossain et~al.}(2008)\citenamefont{Hossain,
  Mottershead, Fournier, Bostwick, McChesney, Rotenberg, Liang, Hardy,
  Sawatzky, Elfimov et~al.}}]{Hossain:2008}
\bibinfo{author}{\bibfnamefont{M.~A.} \bibnamefont{Hossain}},
  \bibinfo{author}{\bibfnamefont{J.~D.~F.} \bibnamefont{Mottershead}},
  \bibinfo{author}{\bibfnamefont{D.}~\bibnamefont{Fournier}},
  \bibinfo{author}{\bibfnamefont{A.}~\bibnamefont{Bostwick}},
  \bibinfo{author}{\bibfnamefont{J.~L.} \bibnamefont{McChesney}},
  \bibinfo{author}{\bibfnamefont{E.}~\bibnamefont{Rotenberg}},
  \bibinfo{author}{\bibfnamefont{R.}~\bibnamefont{Liang}},
  \bibinfo{author}{\bibfnamefont{W.~N.} \bibnamefont{Hardy}},
  \bibinfo{author}{\bibfnamefont{G.~A.} \bibnamefont{Sawatzky}},
  \bibinfo{author}{\bibfnamefont{I.~S.} \bibnamefont{Elfimov}},
  \bibnamefont{et~al.}, \bibinfo{journal}{Nat Phys}
  \textbf{\bibinfo{volume}{4}}, \bibinfo{pages}{527} (\bibinfo{year}{2008}).

\bibitem[{\citenamefont{Chang et~al.}(2008)\citenamefont{Chang, Sassa,
  Guerrero, Mansson, Shi, Pailhes, Bendounan, Mottl, Claesson, Tjernberg
  et~al.}}]{Chang:2008}
\bibinfo{author}{\bibfnamefont{J.}~\bibnamefont{Chang}},
  \bibinfo{author}{\bibfnamefont{Y.}~\bibnamefont{Sassa}},
  \bibinfo{author}{\bibfnamefont{S.}~\bibnamefont{Guerrero}},
  \bibinfo{author}{\bibfnamefont{M.}~\bibnamefont{Mansson}},
  \bibinfo{author}{\bibfnamefont{M.}~\bibnamefont{Shi}},
  \bibinfo{author}{\bibfnamefont{S.}~\bibnamefont{Pailhes}},
  \bibinfo{author}{\bibfnamefont{A.}~\bibnamefont{Bendounan}},
  \bibinfo{author}{\bibfnamefont{R.}~\bibnamefont{Mottl}},
  \bibinfo{author}{\bibfnamefont{T.}~\bibnamefont{Claesson}},
  \bibinfo{author}{\bibfnamefont{O.}~\bibnamefont{Tjernberg}},
  \bibnamefont{et~al.} (\bibinfo{year}{2008}),
  \urlprefix\url{http://arxiv.org/abs/0805.0302}.

\bibitem[{\citenamefont{Chakravarty et~al.}(2003)\citenamefont{Chakravarty,
  Nayak, and Tewari}}]{Chakravarty:2003}
\bibinfo{author}{\bibfnamefont{S.}~\bibnamefont{Chakravarty}},
  \bibinfo{author}{\bibfnamefont{C.}~\bibnamefont{Nayak}}, \bibnamefont{and}
  \bibinfo{author}{\bibfnamefont{S.}~\bibnamefont{Tewari}},
  \bibinfo{journal}{Physical Review B} \textbf{\bibinfo{volume}{68}},
  \bibinfo{pages}{100504} (\bibinfo{year}{2003}).

\bibitem[{\citenamefont{Armitage et~al.}(2001)\citenamefont{Armitage, Lu, Kim,
  Damascelli, Shen, Ronning, Feng, Bogdanov, Shen, Onose
  et~al.}}]{Armitage:2001}
\bibinfo{author}{\bibfnamefont{N.~P.} \bibnamefont{Armitage}},
  \bibinfo{author}{\bibfnamefont{D.~H.} \bibnamefont{Lu}},
  \bibinfo{author}{\bibfnamefont{C.}~\bibnamefont{Kim}},
  \bibinfo{author}{\bibfnamefont{A.}~\bibnamefont{Damascelli}},
  \bibinfo{author}{\bibfnamefont{K.~M.} \bibnamefont{Shen}},
  \bibinfo{author}{\bibfnamefont{F.}~\bibnamefont{Ronning}},
  \bibinfo{author}{\bibfnamefont{D.~L.} \bibnamefont{Feng}},
  \bibinfo{author}{\bibfnamefont{P.}~\bibnamefont{Bogdanov}},
  \bibinfo{author}{\bibfnamefont{Z.~X.} \bibnamefont{Shen}},
  \bibinfo{author}{\bibfnamefont{Y.}~\bibnamefont{Onose}},
  \bibnamefont{et~al.}, \bibinfo{journal}{Physical Review Letters}
  \textbf{\bibinfo{volume}{87}}, \bibinfo{pages}{147003}
  (\bibinfo{year}{2001}).

\bibitem[{\citenamefont{Hussey}(2008)}]{Hussey:2008}
\bibinfo{author}{\bibfnamefont{N.~E.} \bibnamefont{Hussey}}
  (\bibinfo{year}{2008}), \bibinfo{note}{personal communications.}

\bibitem[{\citenamefont{Chen et~al.}(2008)\citenamefont{Chen, Yang, Rice, and
  Zhang}}]{Chen:2008}
\bibinfo{author}{\bibfnamefont{W.~Q.} \bibnamefont{Chen}},
  \bibinfo{author}{\bibfnamefont{K.~Y.} \bibnamefont{Yang}},
  \bibinfo{author}{\bibfnamefont{T.~M.} \bibnamefont{Rice}}, \bibnamefont{and}
  \bibinfo{author}{\bibfnamefont{F.~C.} \bibnamefont{Zhang}},
  \bibinfo{journal}{Euophys. Lett.} \textbf{\bibinfo{volume}{82}}
  (\bibinfo{year}{2008}).

\end{thebibliography}

\end{document}